\documentclass[aps,preprint,superscriptaddress,nofootinbib,preprintnumbers,prd,eqsecnum]{revtex4-1}

\usepackage{bm,amsmath,amsfonts,amssymb,slashed,graphicx,xspace,placeins,rotating,physics,xparse,mathtools}
\usepackage{xcolor,colortbl}
\usepackage{multirow,array,makecell}
\usepackage[inline]{enumitem}
\usepackage{subcaption}
\usepackage{ragged2e}
\usepackage{soul}

\allowdisplaybreaks
\DeclareMathAlphabet{\mathpzc}{OT1}{pzc}{m}{it}

\definecolor{nicered}{rgb}{0.7,0.1,0.1}
\definecolor{nicegreen}{rgb}{0.1,0.5,0.1}
\definecolor{niceblue}{rgb}{0.1,0.1,0.5}
\usepackage[bookmarks=false]{hyperref}
\hypersetup{colorlinks,citecolor= nicegreen,linkcolor= niceblue,urlcolor= niceblue}
\usepackage{breakurl}
\usepackage[force]{feynmp-auto}
\usepackage{psfrag}

\usepackage[capitalise]{cleveref}

\newsavebox{\bigimage}

\newcommand{\nn}{\nonumber}
\newcommand{\GeV}{\text{GeV}\xspace}
\newcommand{\MeV}{\text{MeV}\xspace}

\newcommand{\hhchpt}{\text{HH$\chi$PT}\xspace}

\newcommand{\pln}{p_{\ell\nu}}

\newcommand{\genbar}[1]{\,\overline{\!#1}{}}

\newcommand{\Dx}{D^{(*)}}
\newcommand{\Dss}{D^{**}}
\newcommand{\Bx}{B^{(*)}}

\newcommand{\Bbar}{\genbar{B}}

\newcommand{\mbc}{m_{c,b}}

\newcommand{\tildeM}{\mathbb{M}}

\newcommand{\Api}{\mathbb{A}}
\newcommand{\Vpi}{\mathbb{V}}

\NewDocumentCommand \gXpi {O{0} o m} { 
\IfNoValueTF {#2}
{g^{(#1)}_{#3}}
{g^{(#1)}_{#3,\, #2}}
}

\newcommand{\kpst}{\vb{k}_\star}

\newcommand{\iu}{{\rm i} }
\newcommand{\e}{{\rm e} }
\usepackage{comment}

\newcommand{\amp}[3]{\frac{\langle #1(p') |\, #2\, | #3(p) \rangle}{\sqrt{m_{\let\overline\relax#3} m_{#1}}}}

\makeatletter
\g@addto@macro\bfseries{\boldmath}
\makeatother

\makeatletter
\let\tmp@footnote\footnote
\renewcommand{\footnote}[1]{\tmp@footnote{\linespread{0.9}\selectfont{}#1}}
\makeatother

\makeatletter
\let\temp@caption\caption
\renewcommand{\caption}[2][]{\temp@caption[#1]{\linespread{1.2}\selectfont{}#2}}
\makeatother

\allowdisplaybreaks
\begin{document}

\preprint{CALT-TH-2024-002}
\title{Off-shell vertices in heavy particle  \\ effective theories
and $B\rightarrow D\pi \ell \nu$ }

\author{Michele Papucci}
\affiliation{Walter Burke Institute for Theoretical Physics, California Institute of Technology, Pasadena, CA 91125, USA}

\author{Ryan Plestid}
\affiliation{Walter Burke Institute for Theoretical Physics, California Institute of Technology, Pasadena, CA 91125, USA}

\begin{abstract}
We study the modifications to decay amplitudes in heavy to heavy semileptonic decays with multiple hadrons in the final state due to intermediate heavy hadrons being off-shell or having a finite width.  
Combining Heavy Hadron Chiral Perturbation Theory (\hhchpt) with a BCFW on-shell factorization formula, we show that these effects induce $\mathcal{O}(1/M)$ corrections to the standard results computed in the narrow-width approximation and therefore are important in extracting form factors from data. A combination of perturbative unitarity, analyticity, and reparameterization invariance fully determine these corrections in terms of known Isgur-Wise functions without the need to introduce new form factors. In doing so, we develop a novel technique to compute the boundary term at complex infinity in the BCFW formula for theories with derivatively coupled scalars. While we have used the $\bar B\rightarrow D\pi \ell\nu$ decay as an example, these techniques can generally be applied to effective field theories with (multiple) distinct reference vectors.
\end{abstract}

\maketitle

\twocolumngrid
{
\fontsize{10}{8}\selectfont 
\columnsep20pt
	\tableofcontents
}
\onecolumngrid

\section{Introduction \label{intro} }
Exclusive semileptonic $B\to \Dx \ell \nu$ decays provide one of the most precise channels to measure the CKM matrix element $|V_{cb}|$ and to test charged lepton flavor universality (LFU) \cite{Proceedings:2003xqt,Alonso:2015sja,Buttazzo:2017ixm,Bernlochner:2017jka,Belle-II:2018jsg}.
Upcoming data from Belle II and LHCb will push the precision even further, such that control of the theoretical predictions at (sub-)percent level of accuracy will be required \cite{Belle-II:2018jsg,LHCbCollaboration:2806113}. This includes predictions not only for $B\to \Dx \ell \nu$, but also for related channels involving final-state pions.

Such stringent requirements demand that previously neglected physical effects be included in the analysis. In this paper, we will focus on corrections that stem from intermediate heavy particles (either $B$ or $\Dx$) being pushed slightly off-shell by a soft pion. This can modify the weak-current's matrix element at sub-leading order and must be included for percent-level precision. Our results can be applied to other processes involving heavy particles and additional pions ($B \rightarrow D^{(*)}\ell\nu \pi\pi$) or soft-final state radiation (e.g., $pp \rightarrow pp \gamma$ or $B\to \Dx \ell \nu \gamma$).

In what follows, we show, using a Britto-Cachazo-Feng-Witten (BCFW)-like construction~\cite{Britto:2005fq}, that ``off-shell corrections'' are captured by evaluating the weak current's matrix element at shifted kinematics. Instead of evaluating at $w=v\cdot v'$ one evaluates at $\tilde{w}= \tilde{v}\cdot v'$ or $\tilde{w}' = v \cdot \tilde{v}'$ where $\tilde{v}^{(\prime)}_\mu$ are slightly shifted to account for the momentum of the emitted pion.

Semileptonic decays of a $B$ meson into an on-shell charmed hadron have been computed in the framework of Heavy Quark Effective Theory (HQET) as a simultaneous expansion in the strong coupling constant and heavy (charm and bottom) quark masses \cite{Georgi:1990um,Manohar:2000dt}. The heavy quark spin flavor symmetry of HQET groups heavy hadrons into multiplets and organizes and relates form factors into calculable pieces and non-perturbative Isgur-Wise (IW) functions \cite{Isgur:1989vq,Isgur:1990yhj,Neubert:1993mb}. The IW functions can be measured experimentally or extracted from lattice QCD (LQCD) calculations. In particular, in the mesonic sector, expressions at next-to-leading heavy quark (HQ) power corrections ($1/\mbc$) are known for $B \to \Dss$ transitions and even at next-to-next-to-leading power ($1/\mbc^2$) for\footnote{These results are only useful in combination with some additional truncation scheme as the proliferation of the IW functions renders HQET non-predictive at second order in the HQ expansion.} $B \to \Dx$ \cite{Bernlochner:2017jka,Bernlochner:2022ywh}.  

This is, however, still not sufficient for the accuracy required by experiments: a full description of semileptonic B decays into $\Dx$ states in association with extra pions, $B \to \Dx \ell \nu + \pi, \pi \pi$, is required at $\order{1/\mbc}$ \cite{Colangelo:2018cnj}. This includes contributions from off-shell resonances in addition to on-shell $\Dss$ decays, already included in the stable limit. Furthermore, even for the $\Dss$ contributions near the resonance peak, a better lineshape description, going beyond the narrow width approximation (NWA), is necessary: some of the $\Dss$, namely the $D_0^*$ and $D_1^*$,  have large widths and there are also indications that this fact may be partially due to the presence of two nearby resonance poles for each state~\cite{Meissner:2020khl, Albaladejo:2016lbb, Guo:2006fu, Guo:2006rp}.

To achieve this goal, the calculation of matrix elements of weak current operators between multi-particle hadronic states at first order in the HQ expansion is needed. Up to now, HQET has only been used to compute matrix elements between one-particle states. One could embark on formulating a theory for HQET matrix elements between multi-particle states. Alternatively, one could discuss transition matrix elements between multiparticle states in full QCD, discuss their factorization properties, and then match them onto HQET, an approach recently developed in~\cite{Manzari:2024nxr}.
Instead, to bypass the notoriously hard questions related to describing exclusive multi-particle states in full QCD,
 we choose to tackle the problem by breaking it into two steps. We first match HQET matrix elements of the weak current onto \hhchpt \cite{Burdman:1992gh,Yan:1992gz,Wise:1992hn,Cho:1992cf,Cho:1992gg,Falk:1992cx,Goity:1994xn,Boyd:1994pa,Boyd:1995pq}, a theory whose degrees of freedom are heavy hadrons (whose large momentum modes have been integrated out) coupled to dynamic light mesons described by the chiral Lagrangian. It provides the most natural framework for studying any effect due to the off-shellness of heavy hadrons and their finite decay widths. Then, we compute pion emission amplitudes using \hhchpt. Since the region of interest for the extra emitted pions is such that $p_\pi \lesssim 400 ~\MeV \ll \Lambda_\chi \sim 1\GeV$ (with $\Lambda_\chi$ the \hhchpt UV cutoff), they are in the regime of validity of the theory. 
 Computing the process $B\to D\ell \nu\pi$ in the infinite mass limit was one of the first applications of \hhchpt~\cite{Boyd:1994pa,Boyd:1995pq}.

\begin{figure}
\sbox{\bigimage}{%
  \begin{subfigure}[b]{.55\textwidth}
  \centering
  \includegraphics[width=\textwidth]{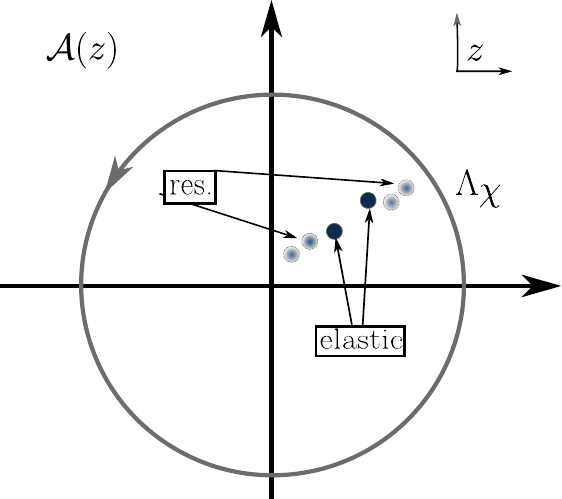}
  \caption{Analytic structure of amplitude}

  \vspace{0pt}
  \end{subfigure}%
}

\usebox{\bigimage}\hfill
\begin{minipage}[b][\ht\bigimage][s]{.4\textwidth}
  \begin{subfigure}{\textwidth}
  \centering
  \includegraphics[width=\textwidth]{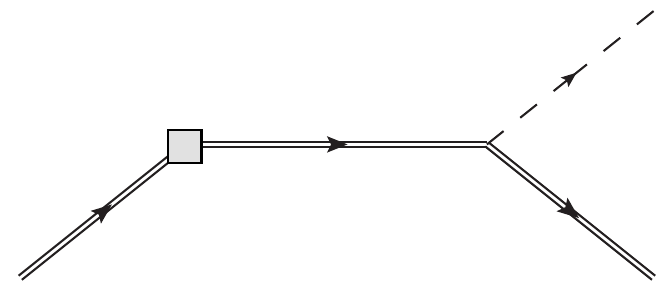}
  \caption{Pion emission off $D$-meson}
  \end{subfigure}%
  \vfill
  \begin{subfigure}{\textwidth}
  \centering
  \includegraphics[width=\textwidth]{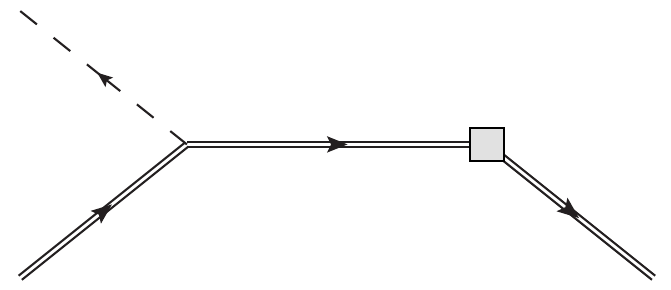}
  \caption{Pion emission off $B$-meson}
  \end{subfigure}

\end{minipage}
\vspace{0pt}
\caption{\justifying In this paper, we consider off-shell amplitudes from the emission of soft particles. These may occur via emission off the initial particle (\textbf{Bottom-Right}) or off the final particle (\textbf{Top-Right}). The weak current is represented with a grey box and the pion with a dashed line. The double lines are heavy particles. We use the analytic structure of the amplitude $\mathcal{A}(z)$ (\textbf{Left}) deformed via a BCFW-like prescription with complex momenta $\pln$ and $p_\pi$ to construct off-shell corrections to the weak current vertex. The cut-off of \hhchpt is denoted schematically as a circle of radius $\Lambda_\chi$. The amplitude, when written as a function of the complex number $z$, contains isolated poles. We use the complex mass scheme for resonances \cite{Denner:2006ic,Actis:2008uh,Denner:2014zga}, which then appear as isolated complex poles at leading order in \hhchpt. \label{summary-figure} }
\end{figure}

Given the precision required, we need to go beyond the leading order in the heavy mass power expansion and track $\order{1/M}$ effects such as $\order{p_\pi/M}$ and $\order{\Gamma/M}$ corrections.
In \hhchpt, power corrections originate from two different sources: $1/2M_{H_{c,b}}$ higher order operators from integrating out the large momentum components of the heavy hadron fields, and $1/2\mbc$ corrections through ``UV'' matching the \hhchpt matrix elements of weak current operators onto the corresponding HQET quantities. The purpose of this paper is to describe in detail how $p_\pi/M$ and $\Gamma/M$ terms originate from these two sources in generality and illustrate it explicitly in a few examples. We will provide the calculations of the full $B\to \Dx \ell \nu + \pi, \pi \pi$, including the relevant $\order{1/\Lambda_\chi}$ ``continuum'' corrections amplitudes elsewhere~\cite{BDpilnu}.

To solve this problem, we adopt the following strategy (see \cref{summary-figure}):
\begin{itemize}
\item Work within \hhchpt assuming analyticity of the amplitudes as functions of the external momenta (inherited from the analytic properties of the parent Lorentz-invariant theory) and perturbative unitarity within the regime of validity of the effective theory.
\item Use a BCFW-inspired factorization formula to decompose multi-particle amplitudes in terms of on-shell (``on-pole'' for the case of unstable resonances) amplitudes with fewer external legs. In particular, this allows matrix elements of the weak current operators to be taken only between heavy hadron one-particle states (albeit with generically non-zero residual momenta $k$, defined by $p_H = M_H v + k$, with $M_H$ the hadron mass and $v$ its velocity used to define \hhchpt).
\item Use reparameterization invariance (RPI) of \hhchpt to relate the one-particle on-shell weak matrix elements at finite residual momenta to those computed at zero residual external momenta.
\item Match the on-shell weak current operator matrix elements between one-particle heavy hadron states onto the corresponding HQET ones (which have been commonly computed at zero residual momenta for the external states).
\end{itemize}
This procedure can describe the full $\Dx\pi$ ($\Dx\pi\pi$) invariant mass spectrum (Dalitz plot). It can be systematically improved by adding higher-order power corrections and/or pion loop corrections in \hhchpt. 

The rest of the paper is organized as follows. In~\cref{NWA}, we will describe the procedure outlined above for the case of pion emissions mediated by (effectively) stable particles whose width can be neglected as is relevant for the $\Dx$. Next, in \cref{FWA}, we consider the necessary generalization to unstable particles. We work in the complex mass scheme and provide a sketch of what would be required to incorporate a broad resonance with a non-trivial lineshape. Explicit calculations for the $D_0^*$, $D_1^*$ states will be provided in~\cite{BDpilnu}. We encode the findings of this paper in a concise set of modified Feynman rules that can be used in \hhchpt calculations to reproduce the results of the BCFW analysis in \cref{sec:Feynman}. Finally, we conclude with a discussion and outlook in \cref{Concl}. 

A series of Appendices complement the main text with various details. In \cref{app:HHchiPT-defs}, we define our conventions for \hhchpt. Next in \cref{app:RPI}, we discuss RPI as a symmetry (redundancy) of \hhchpt, introduce RPI building blocks, and comment on the interplay between vertices and propagators under an RPI transformation. Next, in \cref{app:pion-mels}, we provide explicit RPI pion matrix elements using the aforementioned building blocks. \Cref{app:explicit-soln} provides details of our BCFW construction, discussing an explicit solution for the deformed momenta, the uniqueness of that solution, and a construction of polarization vectors. In \cref{app:alt-def}, we motivate the complex deformation utilized in this paper and compare it to other possible choices. Finally, \cref{app:infinity} illustrates a strategy for applying a BCFW-like construction to derivatively coupled effective field theories with a non-zero contribution from the ``pole at infinity''.

\section{Narrow width analysis \label{NWA}}

To begin, we will analyze the weak coupling limit in which the width of any off-shell intermediate states either vanishes or can be entirely neglected. In full generality, we allow for the off-shell state to have a mass gap relative to the initial or final asymptotic states, which we denote via $\Delta M$ and $\Delta M'$, respectively.  Concretely, all the heavy hadron momenta will be decomposed as
\begin{equation}
    p^{(\prime)}_\mu = M^{(\prime)} v^{(\prime)}_\mu + k^{(')}_\mu~, 
\end{equation}
where primed (non-primed) quantities refer to charm (bottom) quark containing hadrons, with $M^{(\prime)}$ the masses of the lowest-lying hadrons, $M=M_B$, $M'=M_D$. In this way, the residual mass~\cite{Falk:1992fm}, denoted by $\Delta M^{(')}$, will be the mass splitting between the ground and excited hadrons. We will assume that all of the considered intermediate states lie within the range of validity of \hhchpt, i.e., that $\Delta M^{(')} \lesssim \Lambda_\chi$.

The matrix elements for semileptonic decays can be written as 
\begin{equation}
    (2\pi)^4 \delta^{(4)}(p_{\ell\nu}+p_f-p_i)\mathcal{M}  = -\sqrt{8} G_F \int \dd^4 x  ~\e^{\iu p_{\ell\nu} \cdot x }  L_{\Gamma'} \otimes \mel{f}{ \hat{J}_\Gamma (x)}{i}~.
\end{equation}
Upon translating the current to the origin, an overall energy and momentum-conserving delta function is obtained. To simplify the analysis, we factorize the weak vertex and consider the matrix element of the weak current, $\hat{J}_\Gamma$, between an initial hadronic state $\ket{i_v}$ and a final state $\ket{f_{v'}}$, 
\begin{equation}
    \mel{f}{\hat{ J}_\Gamma }{i}_{\rm QCD}= \sqrt{m_B m_D}\int \dd^4 x ~\e^{\iu p_{\ell\nu} \cdot x } \mel{f_{v'}}{\hat{J}_\Gamma(v,v',x)}{i_v}_{\rm \hhchpt}~.
\end{equation}
The initial and final states are labeled by their heavy particles' four velocities and normalized using the conventions in \cref{app:HHchiPT-defs}. For our applications, the initial state will contain a $B$-meson, $\ket{i_v}=\ket{\bar{B}_v}$, and the final state a $D$-meson, $\ket{f_{v'}}=\ket{D_{v'}}$, or a $D$-meson with $n$ soft pions\footnote{In the rest of the paper for ``soft pion'' we mean that $v^{(')} \cdot p_\pi/\Lambda_\chi \ll 1$.}, i.e., $\ket{f_{v'}}=\ket{D_{v'} \pi}$ or $\ket{f_{v'}} = \ket{D_{v'} \pi\pi}$. Our goal is to understand how unitarity, analyticity, and RPI determine the sub-leading corrections to the matrix element in \hhchpt. We formulate our discussion in terms of \hhchpt amplitudes $\mathcal{A}(z)$, which depend on a complex parameter $z$ to be discussed below; the physical amplitudes correspond to $\mathcal{A}(z=0)$. We define the deformed amplitude, $\mathcal{A}(z)$, in \hhchpt between a final state $\ket{f}$ (containing a heavy particle $H'_{v'}$) and an initial state $\ket{i}$ (containing a heavy particle $H_v$) to be given by,
\begin{equation}
\label{eq:J-weak}
    \mathcal{A}(z)\equiv\int \dd^4 x ~\e^{\iu \pln(z) \cdot x } \mel{f_{v'}(z)}{\hat{J}_\Gamma(v,v',x)}{i_v(z)}_{\rm \hhchpt}~.
\end{equation}
The states $\ket{i_v(z)}$ and $\ket{f_{v'}(z)}$ depend on $z$ via kinematic variables.

As described in \cref{intro}, our strategy is to factorize the amplitude into on-shell vertices using a complex deformation of external momenta, like the one used in the BCFW recursion relations \cite{Britto:2005fq}. We choose to deform the momenta of the lepton neutrino pair  and the pion (see \cref{app:alt-def} for a discussion of alternative deformations) via
\begin{equation}
\label{eq:deformation}
    \pln \rightarrow \pln + z q, \quad p_\pi \rightarrow p_\pi - z q~,
\end{equation}
where $z\in \mathbb{C}$ and
\begin{equation}
\label{eq:deform-conditions}
\pln \cdot  q =0, \quad p_\pi \cdot q =0, \quad q^2=0~.
\end{equation}
This choice conserves the total momentum and preserves the mass-shell conditions $\pln^2=m_{\ell\nu}^2$ and $p_\pi^2=m_\pi^2$. By using momentum conservation, one can also show that 
\begin{align}
    \label{ratio-identity}
    \frac{v\cdot q}{v'\cdot q}= \frac{M'}{M}~ .
\end{align}
An explicit expression for $q^\mu$ respecting these constraints is given in \cref{app:explicit-soln}.  We find that the $q^\mu$ components satisfy quadratic equations and that there are two unique solutions\footnote{These solutions correspond physically to a convention labeling spin-up vs. spin-down along a particular quantization axis. See \cref{app:pol-vecs} for more details.} up to a rescaling $q \rightarrow \mathfrak{z}q$ with $\mathfrak{z} \in \mathbb{C}$. Due to the finite masses of the pion and $\ell\nu$ pair, the vector $q^{\mu}$ will always have at least two complex components. 
Deforming the lepton pair and pion momenta allows us to keep the heavy hadron velocities untouched and the external residual momenta vanishing. This is convenient for subsequently matching the weak current amplitude onto known HQET calculations.

In general, under this deformation, the leptonic weak current matrix element may acquire a $z$-dependence. However, because the neutrino is left-handed and massless, for each charged lepton polarization and leptonic current Dirac structure, one can always find a deformation of the lepton and neutrino momenta such that \cref{eq:deformation,eq:deform-conditions} are satisfied and the leptonic weak matrix element is $z$-independent, as is shown in \cref{app:lnu-deform}.
Therefore, we will not discuss the leptonic matrix element any further in what follows.

Following the BCFW construction \cite{Britto:2005fq}, the function to which the Cauchy theorem will be applied is $\mathcal{A}(z)/z$. The physical amplitude can then be obtained as the residue at $z=0$,
\begin{equation}
\mathcal{A}_{\rm phys} = \mathcal{A}(0)=\frac{1}{2\pi \iu } \oint_{\gamma_\epsilon} \dd z  \frac{1}{z} \mathcal{A}(z).
\end{equation} 
where, as usual, the contour $\gamma_\epsilon$ can be taken as a small circle of radius $\epsilon$ around the $z$ origin.

Since \hhchpt is an EFT with a finite cutoff $\Lambda_\chi$ and therefore not BCFW-constructible~\cite{Cheung:2015cba}, we will always be left with a non-vanishing integral at a large radius of $R_\infty \sim |\Lambda_\chi/v^{(\prime)} \cdot q|$ where $\Lambda_\chi$ is the cutoff of our theory. The so-called boundary term originating from this ``pole at infinity'' will contain the contribution of higher-dimensional contact operators~\cite{Feng:2009ei,Cohen:2010mi}. Interestingly, we find the boundary term also subtracts certain unphysical contributions that arise from derivatively coupled pions, as is discussed in detail in \cref{sec:pole-infinity}. We provide a constructive prescription to compute the boundary term explicitly in terms of these higher-dimensional contact operators.  The coefficients of these higher dimensional operators, suppressed by powers of $\Lambda_\chi$ which parameterize the residual ignorance of the ``continuum'' contribution\footnote{These continuum contributions also include the effects of resonances lying outside the regime of validity of \hhchpt, such as the so-called ``sub-threshold'' bottom-charm resonances~\cite{Boyd:1997kz} whose poles are at $z \gg R_\infty$.} in $B\to \Dx \ell \nu \pi$, must be fixed empirically from their contributions to physical observables or via a matching calculation if it exists (e.g., from lattice QCD).

Since we work at tree level, branch cuts are absent and $\mathcal{A}(z)$ only has poles when some intermediate heavy hadron line goes on-shell as depicted in \cref{summary-figure}. Those poles will be inside the large radius circle, $z<R_\infty$, i.e., within the radius of validity of \hhchpt. 
Within \hhchpt, the poles will occur when $v\cdot k - \Delta M \rightarrow 0$ or $v'\cdot k - \Delta M'\rightarrow0$, where $k$ is the residual momentum of the internal line, a function of $p_\pi(z)$ and $\pln (z)$. 
When the amplitude is expanded order-by-order in $1/M$, the analytic structure of $\mathcal{A}(z)/z$ will generically contain higher-order (e.g., double) poles, as shown in the following. 

In the EFT language, these can be identified as originating from insertions of higher power two-point operators, such as the kinetic energy operator on the intermediate heavy hadron line. Working at next-to-leading power, $\order{1/M^{(')}}$, the generic analytic structure contains at most second-order poles. The amplitude can be written as, 
\begin{equation}
     \mathcal{A}_{\rm phys} = \mathcal{A}(0)= ~- \!\!\!\!\!\sum_{{\rm poles},\, {\rm poles}^\prime}  \Bigg\{ \bigg[{\rm Res}_1~\frac{\mathcal{A}(z)}{z} \bigg]_{z_\star} + \bigg[{\rm Res}_2~\frac{\mathcal{A}(z)}{z}\bigg]_{z_\star} \Bigg\} +  \frac{1}{2\pi \iu} \oint_\gamma {\dd z}\frac{{\mathcal{A} }(z)}{z},
    \label{eq:cauchy}
\end{equation}
where $\gamma$ winds counter-clockwise, the residues are taken near the poles, $z=z_\star$ and with the notation ``poles$^{(\prime)}$''  we have made explicit that the sum includes both poles on a charm- or a bottom-containing intermediate heavy hadron line. The contour $\gamma$ is taken such that $|z| \lesssim R_\infty$, while 
\begin{equation}
     {\rm Res}_n~f(z)= \lim_{z\rightarrow z_\star} \frac{1}{(n-1)!}\bigg(\dv{z}\bigg)^{n-1}f(z)~.
\end{equation}
Explicitly we have ${\rm Res}_1 (\mathcal{A}(z)/z)  = \mathcal{A}(z_\star)/z_\star$ and  ${\rm Res}_2 (\mathcal{A}(z)/z)  = -\mathcal{A}'(z_\star)/z_\star^2$.

On each pole, $z=z^{(\prime)}_\star$, an intermediate particle goes on-shell, and the corresponding residue factorizes into left and right on-shell amplitudes. We can use momentum conservation of these sub-amplitudes to determine the position of the pole. Using the pion emission amplitude to fix the values of $z^{(\prime)}_\star$ and working at $\order{1/M^{(\prime)}}$ one gets\footnote{Differently than the on-shell amplitude literature conventions of using all momenta either incoming or outgoing, here we keep the momenta convention determined by the decay kinematics.}
\begin{align}
    \label{zstar-noprime}
& -v \cdot (p_\pi - z q) -\Delta M + \frac{ m_\pi^2- (\Delta M)^2}{2 M} = 0 \nn \\
 & \hspace{0.1\linewidth}\implies  z_\star = \frac{v \cdot p_\pi +\Delta M}{v \cdot q} - \frac{m_\pi^2- (\Delta M)^2}{2 M v \cdot q} ~, \\
& v' \cdot ( p_\pi - z  q) -\Delta M' + \frac{ m_\pi^2- (\Delta M')^2}{2 M'} = 0 \nn \\ 
 & \hspace{0.1\linewidth} \implies  z'_\star = \frac{v' \cdot p_\pi -\Delta M'}{v' \cdot q} + \frac{m_\pi^2- (\Delta M')^2}{2 M' v' \cdot q}  ~,
 \label{zstar-prime}
\end{align}
where $z_\star$ ($z'_\star$) corresponds to a factorization channel where the pion is emitted by the bottom (charm) hadron. Using \cref{ratio-identity}, the above expressions, and defining 
\begin{equation}
    k_\star^{(\prime)}\equiv p_\pi - z_\star^{(\prime)} q~.
\end{equation}
In the vicinity of the poles, the amplitude behaves as
\begin{align}
    \label{vicin-pole-noprime}
    \lim_{z\rightarrow z_\star} \frac{\mathcal{A}(z)}{z} &= -\frac{\mathcal{A}^{(s)}_L(z) \mathcal{A}^{(h)}_R(z)}{2 z v \cdot q}\frac{1}{(z-z_\star)}  + \order{(z-z_\star)^0} \nn \\
    &\simeq 
    -\frac{\mathcal{A}^{(s)}_L(z) \mathcal{A}^{(h)}_R(z)}{2z (v \cdot q)} \frac{1}{ (z-z_\star^{(0)})}  +\frac{\mathcal{A}^{(s)}_L(z) \mathcal{A}^{(h)}_R(z)}{2 z (v \cdot q)^2}\frac{1}{(z-z_\star^{(0)})^2}\frac{m_\pi^2- (\Delta M)^2}{2 M} \\
    &\hspace{0.4\linewidth}+ \order{(z-z_\star^{(0)})^0,1/M^2} ~, \nn\\
    \lim_{z\rightarrow z'_\star} \frac{\mathcal{A}(z)}{z} &= \frac{\mathcal{A}^{(h)}_L(z) \mathcal{A}^{(s)}_R(z)}{2 z v' \cdot q}\frac{1}{(z-z'_\star)}  + \order{(z-z_\star')^0} \nn \\
    &\simeq 
    \frac{\mathcal{A}^{(h)}_L(z) \mathcal{A}^{(s)}_R(z)}{2 z (v' \cdot q)} \frac{1}{ (z-z_\star^{\prime(0)})}  +\frac{\mathcal{A}^{(h)}_L(z) \mathcal{A}^{(s)}_R(z)}{2 z (v' \cdot q)^2}\frac{1}{(z-z_\star^{\prime(0)})^2}\frac{m_\pi^2- (\Delta M')^2}{2 M'}  \label{vicin-pole-prime} \\
    &\hspace{0.4\linewidth}+ \order{(z-z_\star^{(0)\prime})^0,1/M'^2}~,\nn
\end{align}
where  $z_\star^{(0)} = (v \cdot p_\pi +\Delta M) / v \cdot q$, and $z_\star^{\prime(0)} = (v' \cdot p_\pi -\Delta M') / v' \cdot q$, are the $z^{(\prime)}_\star$ solutions at leading order in $1/M^{(\prime)}$. Notice that if we weren't to expand $z_*^{(\prime)}$ in powers of $1/M^{(\prime)}$, i.e., had we remained in the relativistic theory, we would have only single poles. Taking that residue and then expanding $z_*^{(\prime)}$ afterward, one would have obtained the same result as considering both single and double pole contributions.

In certain regions of the phase space the charmed resonance pole may migrate sufficiently close to the origin such that  $\frac{(m_\pi^2- (\Delta M')^2)}{2M'(v' \cdot p_\pi - \Delta M')}\sim \order{1}$ (in $B$ decays this only happens for charmed resonances due to kinematics). In this case, the heavy mass power expansion gets slightly modified \cite{Beneke:2003xh,Beneke:2004km,Beneke:2015vfa}. When considering kinematics near a resonance, one should keep the second term in \cref{zstar-prime} when using \cref{vicin-pole-prime} rather than the approximant $z_\star^{\prime(0)}$. 

Since we will not consider such resonant kinematics in what follows, and we work at fixed order in the heavy mass power expansion, we drop the superscript ``$(0)$'' from $z^{(\prime)}_\star$ hereafter. The superscripts $(h)$, $(s)$ on $\mathcal{A}_{L,R}$ indicate that those amplitudes describe a weak $b \to c \ell \nu$ transition (``hard'') or a pion emission (``soft'') respectively.

The double pole in the second term automatically reproduces kinetic insertions from the \hhchpt Lagrangian \cite{Manohar:2000dt}: the residual momentum is $k^2 = p_\pi^2 = m_\pi^2$ and the quantity in the numerator of the double pole terms is   %
\begin{equation}
    \kpst^{(\prime)2}=(k_\star - v^{(\prime)}\cdot k_\star)^2 = m_\pi^2 - (\Delta M^{(\prime)})^2~.
\end{equation}
where the bold face indicates a $v^{(')}$-transverse quantity.
Taking the residues of \cref{vicin-pole-noprime,vicin-pole-prime} one gets 
\begin{align}
{\rm Res}_1~\frac{\mathcal{A}(z)}{z}\Big|_{z=z_\star} & = -\frac{\mathcal{A}^{(s)}_L(z_\star) \mathcal{A}^{(h)}_R(z_\star)}{2z_\star (v \cdot q)}, \label{eq:residue-1-noprime}\\
{\rm Res}_2~\frac{\mathcal{A}(z)}{z}\Big|_{z=z_\star} & = -\frac{\kpst^2}{4 M}\frac{1}{ (z_\star v \cdot q)^2} \Bigg[\mathcal{A}^{(s)}_L(z_\star) \mathcal{A}^{(h)}_R(z_\star) - z_\star  \dv{z} \Big(\mathcal{A}^{(s)}_L(z) \mathcal{A}^{(h)}_R(z)\Big)\Big|_{z=z_\star}\Bigg],\label{eq:residue-2-noprime}\\
{\rm Res}_1~\frac{\mathcal{A}(z)}{z}\Big|_{z=z'_\star} & = \frac{\mathcal{A}^{(h)}_L(z'_\star) \mathcal{A}^{(s)}_R(z'_\star)}{2z'_\star (v' \cdot q)},\label{eq:residue-1-prime}\\
{\rm Res}_2~\frac{\mathcal{A}(z)}{z}\Big|_{z=z'_\star} & = -\frac{\kpst^{\prime 2}}{4 M'}\frac{1}{(z'_\star v' \cdot q)^2}\Bigg[\mathcal{A}^{(h)}_L(z'_\star) \mathcal{A}^{(s)}_R(z'_\star) - z'_\star  \dv{z} \Big(\mathcal{A}^{(h)}_L(z) \mathcal{A}^{(s)}_R(z)\Big)\Big|_{z=z'_\star}\Bigg].\label{eq:residue-2-prime}
\end{align}
The terms $(z^{(')}_\star v^{(')} \cdot q )$ in the previous expressions reproduce the scalar tree level off-shell inverse propagators
\begin{align}
    2(z_\star v \cdot q)   &= 2(v \cdot p_\pi +\Delta M)   = -\iu D^{-1}_{v} (-p_\pi), \\
    2(z'_\star v' \cdot q) &= 2(v' \cdot p_\pi -\Delta M') =  \iu D^{-1}_{v'} (p_\pi).
\end{align}
To recover the full amplitude, one needs to sum over all the intermediate states going on-shell at that pole. These sums reproduce the numerators of \hhchpt propagators.

\subsection{Weak current on-shell amplitude}

The matrix elements $\mathcal{A}^{(h,s)}_{L, R}$ are taken on-shell but contain a complex residual momentum $k^{(\prime)}_\star = k(z^{(\prime)}_\star) \neq 0$; they also need to be consistently expanded at the correct order in the heavy mass. This expansion is greatly aided by RPI transformations.  \hhchpt has a hidden Lorentz invariance that presents itself as an RPI symmetry (redundancy) in the effective theory \cite{Luke:1992cs,Heinonen:2012km}. (see \cref{app:RPI-BB} for a discussion). Therefore, matrix elements with finite residual momentum can be obtained from matrix elements with vanishing external residual momenta by performing an RPI redefinition on the fields and states.

When considering matrix elements with residual momentum $k_\star^{(\prime)}$ we therefore introduce a modified reference vector $\tilde{v}^{(\prime)}$. For the ground state multiplet (with $\Delta M=0$), 
\begin{align}
    \tilde{v}^{(\prime)} &=v^{(\prime)} +\frac{k_\star^{(\prime)}}{M^{(\prime)}}~,  ~\\ 
    \tilde{k}^{(\prime)}&=0~. 
\end{align}
When considering excited states, we must retain non-zero residual momentum, but we can align it parallel with $\tilde{v}^{(\prime)}$ as discussed in \cref{app:RPI}. 

Invariance under RPI transformations then implies, 
\begin{equation}
    \label{eq:NLP-matching}
  \begin{split}
    \mel{H'_{v'}(k'_\star)}{\hat{J}_\Gamma(v',v)}{H_v(0)} &=  \mel{H'_{\tilde{v}'}(0)}{\hat{J}_\Gamma(\tilde{v}',v)}{H_v(0)} + \order{1/ M^{(\prime)2}} ~ \\
      &=  \mel{H'_{\tilde{v}'}(0)}{\hat{J}_\Gamma(\tilde{v}',v)}{H_v(0)}_{1/\mbc^0} ~ \\
      &\hspace{0.1\linewidth} + \mel{H'_{v'}(0)}{\hat{J}_\Gamma(v',v)}{H_v(0)}_{1/\mbc} \\
      &\hspace{0.125\linewidth}+ ~\order{1/\mbc^2, 1/M^{(\prime)2},1/ M^{(\prime)}\mbc}~, 
      \end{split}
\end{equation}
and similarly for the case where the bottom hadron acquires a residual momentum. 
The $\order{1/\mbc^{0,1}}$ subscripts on the matrix elements refer to the order of the heavy quark mass expansion derived from the \hhchpt matching to HQET in the UV~\cite{Boyd:1994pa,Boyd:1995pq}.\!\footnote{
For simplicity's sake, we work at leading order in \hhchpt i.e., \ neglecting chiral loops, such that for the matrix element written above, only the zero-pion contributions to $\hat{J}_\Gamma$ are relevant.} Expanding in $1/M^{(\prime)}$ then gives, 
\begin{align}
    \begin{split}
    \label{vertex-correction-1}
    \mathcal{A}_L^{(h)}(z_\star')&=\mel{H'_{v'}(0)}{\hat{J}_\Gamma}{H_v(0)}_{1/\mbc^0}\\
    &\hspace{0.05\linewidth}+ \frac{k_\star'^\mu}{M'}\pdv{v'^\mu}\mel{H'_{v'}(0)}{\hat{J}_\Gamma}{H_v(0)}_{1/\mbc^0}  \\
    &\hspace{0.05\linewidth} + \mel{H'_{v'}(0)}{\hat{J}_\Gamma}{H_v(0)}_{1/\mbc}  + \mathcal{O}(1/M^{\prime2}) ~,
    \end{split}
    \\
    \begin{split}
    \mathcal{A}_R^{(h)}(z_\star)&=\mel{H'_{v'}(0)}{\hat{J}_\Gamma}{H_v(0)}_{1/\mbc^0}\\
    &\hspace{0.05\linewidth}+ \frac{k_\star^\mu}{M}\pdv{v^\mu}\mel{H'_{v'}(0)}{\hat{J}_\Gamma}{H_v(0)}_{1/\mbc^0}  \\
    &\hspace{0.05\linewidth} +
     \mel{H'_{v'}(0)}{\hat{J}_\Gamma}{H_v(0)}_{1/\mbc} 
    + \mathcal{O}(1/M^2) ~. \label{vertex-correction-2} 
    \end{split}
\end{align}
When taking the derivative with respect to $v^{(\prime)}$, one must also consider polarization vectors. Polarizations implicitly depend on $v^{(\prime)}$ since they must satisfy the transversality condition $v\cdot \epsilon=0$ which implies $\pdv{v^\nu}\epsilon^\mu = - v^\mu \epsilon^\nu$.

In performing the matching, one should consistently use the RPI building blocks discussed in \cref{app:RPI-BB}. Schematically, one promotes heavy fields $H_v \rightarrow \mathcal{H}_v$ to objects that only pick up a phase under an RPI transformation. For example the label $v$ and $v'$ are promoted to $\mathcal{V}$ and $\mathcal{V}'$. Importantly, when acting on states with vanishing residual momentum, both fields yield equivalent matrix elements, i.e., $\mathcal{H}_v \ket{H_v(0)} = H_v \ket{H_v(0)}$. As an explicit example, let us consider the vector current between pseudoscalar states, 
\begin{equation}
  \begin{split}
    \label{simple-example}
    \mel{D_{v'}(k'_\star)}{\hat{J}_V^\mu}{\Bbar_v(0)}_{1/\mbc^0} &=  -
      \xi(\mathcal{V}\cdot \mathcal{V}')\mel{D_{v'}(k'_\star)}{ {\rm Tr}\big[~ \overline{\cal H}'_{v'} \gamma^\mu {\cal H}_v~ \big]}{\Bbar_v(0)}_{1/\mbc^0} \\
    &=  -\xi(\mathcal{V}\cdot \tilde{\mathcal{V}}')
      \mel{D_{\tilde{v}'}(0)}{{\rm Tr}\big[~ \overline{\cal H}'_{\tilde{v}'} \gamma^\mu {\cal H}_v~ \big]}{\Bbar_v(0)}_{1/\mbc^0} \\
      &= -\xi(v\cdot \tilde{v}')
      \mel{D_{\tilde{v}'}(0)}{{\rm Tr}\big[~ \overline{H}'_{\tilde{v}'} \gamma^\mu H_v~ \big]}{\Bbar_v(0)}_{1/\mbc^0} \\
      &= \xi(v\cdot \tilde{v}')[v^\mu + \tilde{v}'^\mu]\\
      &= \xi(v\cdot v')[v^\mu + v'^\mu] + \frac{v\cdot k_\star}{M'} \xi'(v\cdot v')[v^\mu + v'^\mu] 
      \\
    & \hspace{0.1\linewidth}
     + \frac{k_\star^\mu}{M'}  \xi(v\cdot v')~ + \mathcal{O}(1/M^2).
  \end{split}
\end{equation}
The first line defines the current's matrix element in terms of RPI building blocks (see \cref{app:RPI-BB}). The second line follows from the RPI symmetry of \hhchpt. In going from the second to the third line, we have used $\mathcal{H}_v \ket{H_v(0)} = H_v \ket{H_v(0)}$. The fourth line follows algebraically, and the fifth line is an expansion of $\tilde{v}^{\prime}$ in $1/M'$. 

Before proceeding, let us comment on the role of complex reference velocity vectors (note that $k_\star$ contains complex components that are necessary to push intermediate lines on-shell). In the case of \hhchpt, this is a trivial extension of well-known analytic properties of Lorentz invariant amplitudes because RPI is just redundancy of a momentum redefinition, $p = M v + k$. As we will see in \cref{sec:pole-infinity}, these complex components are ultimately subtracted, and only matrix elements with real-valued reference vectors appear in our final formula \cref{Born-example-v2}. Therefore, we only ever need to perform matching calculations from HQET to \hhchpt using real-valued $v$ and $v^\prime$ and never encounter complex reference vectors in HQET calculations like those discussed in \cite{Denner:2014zga}.

\subsection{Pion emission amplitudes}
Momentum conservation dictates that pion emission amplitudes always involve finite residual momenta for at least one heavy particle. 
For consistency at $\order{1/M^{(\prime)}}$ one therefore needs to use the RPI-invariant fields to write the pion emission vertex, irrespective of whether an RPI transformation is performed on $\mathcal{A}^{(s)}_{L,R}$.
Explicitly these soft, $(s)$, matrix elements are defined as 
\begin{align}
    \label{eq:NLP-matching-soft-1}
    \mathcal{A}^{(s)}_{L}(z) &= g_H\mel{\pi(p_\pi) H_v'(-p_\pi)}{ {\rm Tr} \qty[\mathcal{H}_v \gamma_5 \slashed{\Api} \overline{\mathcal{H}}_v]}{H_v(0)}_{1/m_{b,c}^0} + \mathcal{O}(1/m_{b,c}^1) ~,\\
    \mathcal{A}^{(s)}_{R}(z) &= g_H\mel{\pi(p_\pi) H_{v'}'(0)}{ {\rm Tr} \qty[\mathcal{H}_{v'}' \gamma_5 \slashed{\Api} \overline{\mathcal{H}}_{v'}]}{H_{v'}(p_\pi)}_{1/m_{b,c}^0} + \mathcal{O}(1/m_{b,c}^1)~,
     \label{eq:NLP-matching-soft-2}
\end{align}
where the script operators correspond to RPI building blocks discussed in \cref{app:RPI-BB} and the coupling $g_H$ is defined in \cref{eq:lagr-LO}. The $1/m_{b,c}^1$ corrections to the matrix elements can be found in \cite{Boyd:1994pa,Casalbuoni:1996pg}. In the above expression, $p_\pi(z)$ is the deformed pion's momentum. 
As anticipated above, it is, however, convenient to shift the pion emission amplitude so that the same velocity $\tilde v^{(\prime)}$ flows along the internal line. That amounts to swapping the heavy hadron lines having zero and finite residual momentum (see \cref{app:pion-mels}).

Furthermore, as the pions are derivatively coupled,\!\footnote{We neglect here in the discussion higher-order interactions proportional to $m_\pi/\Lambda_\chi$.}  their emission amplitude depends on $z$ at LO in the power expansion. Therefore for the double pole residue at $\order{1/M^{(\prime)},1/\Lambda_\chi}$ one has
\begin{equation} \label{deriv-cancels}
\begin{split}
\Bigg[\mathcal{A}^{(s)}_L(z_\star) \mathcal{A}^{(h)}_R(z_\star) - z_\star  \dv{z} \Big(\mathcal{A}^{(s)}_L(z) \mathcal{A}^{(h)}_R(z)\Big)\Big|_{z=z_\star}\Bigg]  \\
 & \hspace{-0.2\linewidth} = \Bigg[\mathcal{A}^{(s)}_L(z_\star)  - z_\star  \dv{z} \mathcal{A}^{(s)}_L(z) \Big|_{z=z_\star}\Bigg]  \mathcal{A}^{(h)}_R(0)+ \order{1/M} \\
 & \hspace{-0.2\linewidth} =   \mathcal{A}^{(s)}_L(0) \mathcal{A}^{(h)}_R(0) + \order{1/M, 1/\Lambda_\chi} 
\end{split}
\end{equation}
and similarly for the case of the double pole at $z=z'_\star$. The last line holds at $\order{1/\Lambda_\chi}$ and neglecting quark mass effects as $\mathcal{A}^{(s)}_L$ is linear in $ p_\pi$ and vanishes for $ p_\pi \to 0$ at this order. Therefore at $\order{1/M^{(\prime)2},1/\Lambda_\chi}$ the double poles (i.e., \cref{eq:residue-2-noprime,eq:residue-2-prime}) reproduces exactly the Feynman diagram one would have written using leading order vertices and one $1/M$ Lagrangian insertion on the internal line (see \cref{app:RPI-propagator,sec:pole-infinity} for further discussion).

\subsection{Large-$z$ contributions \label{sec:pole-infinity}}

As mentioned earlier, the contribution from the ``pole at infinity'' does not vanish because of derivative couplings in \hhchpt. In a general EFT, the contour at large-$z$ contains terms originating from spurious poles from factorization channels that cannot be probed with a chosen $z$-deformation~\cite{Hodges:2009hk}, terms originating from ``bad'' UV behavior of interaction vertices, and contact terms from higher dimensional operator insertions. In the specific case of the deformation given in \cref{eq:deformation}, there are no spurious pole contributions:  heavy quark number and lepton number conservation (and the requirement that we will only consider $b\to c$ weak transitions) eliminate any other factorization channel not captured by this deformation.
However, since the pion is derivatively coupled in $\chi$PT and \hhchpt, any pion vertex will introduce $z^1$ dependence, which can modify \hhchpt power counting when $z$ is taken to large values. 

We pursue a strategy that is well suited to \hhchpt.  The idea is to express the large-$z$ limit of the amplitude as a tensor contracted with deformed pion momenta, $\hat{p}_\pi(z) = -z q + \mathcal{O}(1)$ where $\hat{p}_\pi=p_\pi -zq$. The tensor is $\mathcal{O}(1)$ as $z\rightarrow \infty$, and can be reconstructed using Cauchy's theorem. Effectively, this amounts to applying the Cauchy theorem to derivatives of $\mathcal{A}(z)$ and is analogous to subtractions used in the context of traditional dispersion relations. We are not aware of this approach being discussed in the literature for the construction of EFT amplitudes using BCFW techniques. 

Let us consider the $n^{\rm th}$ order in the chiral expansion, $\mathcal{A}^{(n)}$ and define a set of $z$-dependent tensors, 
\begin{equation}
    \mathcal{A}^{(k,n)}_{\mu_1\ldots \mu_k} (z)= \frac{\partial^k}{\partial \hat{p}_\pi^{\mu_1}\ldots \partial \hat{p}_\pi^{\mu_k}}
    \mathcal{A}^{(n)}~.
\end{equation}
In the large-$z$ limit, all of the $\mathcal{A}^{(k,n)}_{\mu_1\ldots \mu_k} (z)$ have a dispersive representation,
\begin{equation}\label{eq:disp-tensor}
    \frac{1}{(2\pi\iu)} \oint_{\Gamma} \dd w  \frac{\mathcal{A}^{(k,n)}_{\mu_1\ldots \mu_k} (w)}{w-z} = \mathcal{A}^{(k,n)}_{\mu_1\ldots \mu_k} (z) + \sum {\rm Res}~ \frac{1}{w-z}\mathcal{A}^{(k,n)}_{\mu_1\ldots \mu_k}(w)~.
\end{equation}
where $\Gamma$ is a contour at large-$z$ (not necessarily related to $\gamma$), and the residues are taken with respect to $w$. 
An iterative procedure can be used when multiple such tensors must be constructed; one first constructs the ``most divergent'' piece, then subtracts it, and repeats. After constructing all of the necessary $\mathcal{A}^{(k)}_{\mu_1\ldots \mu_n} (z)$ tensors, the $n^{\rm th}$ order amplitude has an unambiguous dependence on $z$ valid for large values of $z$. We may, therefore, directly evaluate the amplitude along the large-$z$ contour and compute its contribution to the BCFW factorization formula explicitly.
The effects of higher dimensional operators will enter through the left-hand side of \cref{eq:disp-tensor}, while the second term of its right-hand side can be directly computed in terms of factorized amplitudes on poles at finite $z$ discussed earlier. A more detailed discussion, with the explicit calculation of the boundary term for the leading order amplitude and a sketch of the procedure for the sub-leading order one, is given in \cref{app:infinity}.
One finds
\begin{equation}
\frac{1}{(2\pi\iu)} \oint_\gamma \dd{z} \frac{{\mathcal{A}}(z)}{z} = ~ C_\gamma^{(0)} + \frac{1}{\Lambda_\chi} C_\gamma^{(1,{\rm pole})} + \frac{1}{\Lambda_\chi} C_\gamma^{(1,{\rm contact})} + \order{1/M^{(\prime)},1/\Lambda_\chi^2}. \nn
\end{equation}

The constants $C_\gamma^{(0)}$ and $C_\gamma^{(1,{\rm pole})}$ arise from the ``bad'' large-$z$ behavior of derivatively coupled pion interaction terms.\!\footnote{The constant $C_\gamma^{(0)}$ is $\mathcal{O}(1)$ because nominally suppressed contributions of $\mathcal{O}(p_\pi^n/\Lambda_\chi^n)$ become $\mathcal{O}(1)$ when $z\sim \mathcal{O}(\Lambda_\chi/p_\pi)$.}
The constant $C_\gamma^{(0)}$, given in \cref{eq:Cgamma-leading} is effectively a subtraction of certain unphysical contributions from the sum over residues in \cref{eq:cauchy} in the leading order amplitude. A similar expression holds for $C_\gamma^{(1,{\rm pole})}$. At $\mathcal{O}(1/\Lambda_\chi)$, there are contributions from contact terms, $C_\gamma^{(1,{\rm contact})}$, parameterized as the most general set of higher dimensional operators, which read
\begin{equation}
\frac{1} {\Lambda_\chi} C_\gamma^{(1,{\rm contact})} = \begin{multlined}[t]
 \frac{f_{1}(v \cdot v')}{\Lambda_\chi} \mel{D_{v'}(0)}{{\rm Tr}\big[~ \overline{H}'_{v'} \Gamma H_v \gamma^5 \slashed{\Api}~\! (p_\pi) \big]}{\Bbar_v(0)} \\ 
 + \frac{f_2(v \cdot v')}{\Lambda_\chi} \mel{D_{v'}(0)}{{\rm Tr}\big[~ \overline{H}'_{v'} \Gamma H_v \gamma^5 (v+v')\cdot \Api ~\! (p_\pi) \big]}{\Bbar_v(0)} \\
 + \frac{f_3(v \cdot v')}{\Lambda_\chi} \mel{D_{v'}(0)}{{\rm Tr}\big[~ \overline{H}'_{v'} \Gamma H_v \gamma^5 (v-v')\cdot \Api ~\! (p_\pi) \big]}{\Bbar_v(0)}~.
 \end{multlined}\label{eq:higher-dim}
 \end{equation}
 These contributions are physical and can be constrained by a combination of experimental data, lattice QCD, and matching calculations.

\subsection{Full $B \to D \ell \nu \pi$ via intermediate $\Dx$} 
Let us study a concrete example to see how all of the pieces fit together. 
Collecting the various components discussed above and focusing on the case where only $\Dx$ propagate as intermediate states, the amplitude may be written as 
\begin{equation}
    \mathcal{A} = \mathcal{A}_{\rm Born} + \mathcal{A}_{\rm Contact} ~,
\end{equation}
The so-called ``Born graphs'' contain explicit poles in the low-energy theory and constitute our object of interest in the previous subsections. 

Collecting the various terms and summing over the poles of intermediate $\Bx$ and $\Dx$ states, (for simplicity) at leading-order in $1/\Lambda_\chi$, but sub-leading order in $1/M$, our BCFW-like algorithm yields schematically (i.e., suppressing sums over polarizations and leaving implicit a sum over polarization and particles)
\begin{equation}
    \label{Born-example-v1}
    \begin{split}
    \mathcal{A}_{\rm Born}&=  C^{(0)}_\gamma - \qty(\frac{1}{2(z_\star' v'\cdot q)} \mathcal{A}_L^{(h)}(z_\star') \mathcal{A}_R^{(s)}(z_\star') -   \frac{1}{2(z_\star v\cdot q)} \mathcal{A}_L^{(s)}(z_\star) \mathcal{A}_R^{(h)}(z_\star)  )  \\
    &- \qty( - \mathcal{A}_L^{(s)}(0)\frac{1}{2(z_\star v\cdot q)^2} \frac{{\kpst^2}} {2M}   \mathcal{A}_R^{(h)}(0)- \mathcal{A}_L^{(h)}(0)\frac{1}{2(z_\star' v'\cdot q)^2} \frac{{(\kpst')^2}} {2M'} \mathcal{A}_R^{(s)}(0))~.
    \end{split}
\end{equation}
The term $C^{(0)}_\gamma$ subtracts off most of the $z_\star^{(\prime)}$ dependence in the above expression. Only {\it after} this subtraction is the result manifestly independent of $q$, which is related to the fact that one must include all poles (including poles at infinity) to obtain a result which does not depend on $q$. Using \cref{eq:C-gamma-0} we arrive at, 
\begin{equation}
    \label{Born-example-v1p5}
    \begin{split}
    \mathcal{A}_{\rm Born}&=   - \qty(\frac{1}{2(z_\star' v'\cdot q)} \mathcal{A}_L^{(h)}(0) \mathcal{A}_R^{(s)}(0) -   \frac{1}{2(z_\star v\cdot q)} \mathcal{A}_L^{(s)}(0) \mathcal{A}_R^{(h)}(0)  )  \\
    &- \qty( - \mathcal{A}_L^{(s)}(0)\frac{1}{2(z_\star v\cdot q)^2} \frac{{\kpst^2}} {2M} \mathcal{A}_R^{(h)}(0)- \mathcal{A}_L^{(h)}(0)\frac{1}{2(z_\star' v'\cdot q)^2} \frac{{(\kpst')^2}} {2M'} \mathcal{A}_R^{(s)}(0))~.
    \end{split}
\end{equation}
Expanding about $\tilde{v}^{(\prime)}=v^{(\prime)}$, and using $\tilde{v}= v-p_\pi/M$ and $\tilde{v}'=v'+p_\pi/M'$, we obtain
\begin{equation}
    \label{Born-example-v2}
    \begin{split}
    \mathcal{A}_{\rm Born}=& - \qty[\frac{1}{2(z_\star' v'\cdot q)} \mathcal{A}_L^{(h)} \mathcal{A}_R^{(s)} -  \frac{1}{2(z_\star v\cdot q)} \mathcal{A}_L^{(s)} \mathcal{A}_R^{(h)}  ]_{1/m_{b,c}^{(0)}}  \\
    &- \qty[\frac{1}{2(z_\star' v'\cdot q)} \frac{p_\pi^\mu}{M'}\qty(\pdv{v^{\prime\mu}}\mathcal{A}_L^{(h)} \mathcal{A}_R^{(s)}) +   \frac{1}{2(z_\star v\cdot q)}  \frac{p_\pi^\mu}{M}\qty(\pdv{v^{\mu}}\mathcal{A}_L^{(s)}\mathcal{A}_R^{(h)})  ]_{1/m_{b,c}^{(0)}} \\
    &- \qty[ - \mathcal{A}_L^{(s)}\frac{1}{2(z_\star v\cdot q)^2} \frac{{\kpst^2}} {2M}  \mathcal{A}_R^{(h)}- \mathcal{A}_L^{(h)}\frac{1}{2(z_\star' v'\cdot q)^2} \frac{{(\kpst')^2}} {2M'}  \mathcal{A}_R^{(s)}]_{1/m_{b,c}^{(0)}} ~\\
    &- \qty[\frac{1}{2(z_\star' v'\cdot q)} \mathcal{A}_L^{(h)} \mathcal{A}_R^{(s)} -  \frac{1}{2(z_\star v\cdot q)} \mathcal{A}_L^{(s)} \mathcal{A}_R^{(h)}  ]_{1/m_{b,c}^{(1)}}  + ~\ldots ~
    \end{split}
\end{equation}
where all amplitudes are evaluated at $z=0$. 
The first line encodes the result of~\cite{Lee:1992ih}, while the second and third lines are the $1/M^{(\prime)}$ corrections we are after.
The final line includes standard $1/m_{b,c}$ corrections that come from matching HQET onto \hhchpt as discussed explicitly around \cref{vertex-correction-1,vertex-correction-2,eq:NLP-matching-soft-1,eq:NLP-matching-soft-2}.

Notice that without the contribution of $C_\gamma^{(0)}$ the expression for the amplitude would disagree with the results of~\cite{Lee:1992ih}, even in the infinite mass limit.
Before proceedings let us comment on the kinetic energy insertions, which we write as 
\begin{equation}
    \frac{\kpst^2}{2M} = \frac{m_\pi^2-(\Delta M)^2}{2M} = \frac{\vb{p}_\pi^2}{2M} + \frac{[z_\star(v\cdot q) ]^2}{2M}~. 
\end{equation}
The final equality shows that up to contact terms, this expression is equivalent to what one would obtain using conventional heavy-particle Feynman rules. Using $m_\pi^2-(\Delta M)^2$ is convenient because it makes the kinetic energy insertions $z$-independent. If one uses instead $\vb{p}_\pi^2$ then a further contribution arises to $C_\gamma$ which subtracts off an nonphysical contribution from the residue of the pole. Both expressions yield the same physical amplitude. 

Let us write the result for $B\rightarrow D \ell \nu \pi$ summing over all intermediate particles and spins in the ground state multiplet (i.e., ignoring higher resonance contributions). We only need the single-trace contribution from \cref{superfield-propagator} since there is no $DD\pi$ vertex. Let us first keep the full RPI building blocks (see \cref{app:RPI-BB}) without expanding in $\tilde{v}$
\begin{align}
    \begin{split}
    \sum_{\rm states}  \qty[\mathcal{A}_L^{(s)} \mathcal{A}_R^{(h)}]_{1/m_{b,c}^0} &=\\[-3pt]
    &
    -2\xi(v\cdot \tilde{v}')\mel{\pi(p_\pi)D(-p_\pi)}{{\rm Tr}\qty[ \tilde{\Pi}_-'\gamma_5 \slashed{\Api} \bar{\cal H}_{\tilde{v}'}\tilde{\Pi}_+'\Gamma \mathcal{H}_v]}{ \bar{B}(0)}_{1/m_{b,c}^0},\label{explicit-sum-states-1}
    \end{split} 
    \\
    \begin{split}
    \sum_{\rm states}  \qty[\mathcal{A}_L^{(h)} \mathcal{A}_R^{(s)}]_{1/m_{b,c}^0}  &=  \\[-3pt]
    &-2\xi(\tilde{v}\cdot v')\mel{\pi(p_\pi)D(0)}{{\rm Tr}\qty[  \bar{\cal H}_{v'}\Gamma \tilde{\Pi}_+\mathcal{H}_{\tilde{v}} \gamma_5 \slashed{\Api}\tilde{\Pi}_-]}{ \bar{B}(p_\pi)}_{1/m_{b,c}^0}.\label{explicit-sum-states-2}
    \end{split}
\end{align}
We have performed an RPI transformation shifting the residual momentum into the final state $D$-meson, such that the intermediate states have $k=0$. These should be expanded and truncated at $\mathcal{O}(1/M)$ to match the level of precision used in the derivation above. This first order in this expansion produces the second line of \cref{Born-example-v2}. Having performed an RPI transformation so that the internal line has zero residual momentum, we may use the standard \hhchpt propagator for the entire (super) multiplet, i.e., the one given in \cref{superfield-propagator} with the replacement $v\rightarrow \tilde{v}$. We discuss the interplay between vertices and propagators under RPI transformations in \cref{app:RPI-propagator}.

\section{Finite width effects \label{FWA}} 
In the preceding section, we considered the analytic structure of a theory with a spectrum of (approximately) stable particles. Phenomenologically it is interesting to also consider off-shell particles, with a mass splitting $\Delta M$ and a sizeable off-shell width $\Gamma$. 

A convenient method for incorporating a finite width is the complex mass scheme. This scheme has been demonstrated to be gauge invariant and unitary order-by-order in perturbation theory\footnote{Interestingly, the proof of unitarity of the complex mass scheme uses Veltman's largest time equation \cite{Veltman:1963th}, which for the case of QCD at tree level has been shown to be equivalent to a BCFW deformation~\cite{Vaman:2005dt}.}~\cite{Denner:2014zga}. It amounts to treating the renormalized mass as a complex quantity, $M^2 \rightarrow M^2 - \iu \, \Gamma M$,  and to adding suitable complex counter terms whose values are fixed by a scheme-dependent renormalization condition that ensures the analytic structure of the Green's function is maintained (since now a finite width already arises at tree level).  

For our purposes, the complex mass scheme introduces two modifications: 
\begin{itemize}
\item the position of the poles $z_*^{(\prime)}$ in the complex plane are shifted, such that the residual mass shifts (at leading-order in $1/M$) to $\Delta M^{(\prime)} \rightarrow \Delta \tildeM^{(\prime)} = M^{(\prime)}-\iu \Gamma^{(\prime)}/2$,
\item the matching conditions between \hhchpt and HQET are modified via the changes to the relation between the (complex) heavy hadron and (real) heavy quark masses.
\end{itemize}

Focusing first on the shifts of the location of the poles $z_\star^{(\prime)}$ one has 
\begin{align}
    \label{zstar-noprime-gamma}
  z_\star & = \frac{v \cdot p_\pi +\Delta M - \iu \, \Gamma/2}{v \cdot q} - \frac{m_\pi^2- \Delta M (\Delta M - \iu \, \Gamma)}{2 M v \cdot q} \nn \\
              & =  \frac{v \cdot p_\pi +\Delta \tildeM }{v \cdot q} - \frac{m_\pi^2- (\Delta \tildeM)^2 - \Gamma^2 /4}{2 M v \cdot q} ~,\\
  z'_\star & = \frac{v' \cdot p_\pi -\Delta M' + \iu \, \Gamma'/2}{v' \cdot q} + \frac{m_\pi^2- \Delta M' (\Delta M' - \iu \, \Gamma')}{2 M' v' \cdot q} \nn \\
  & = \frac{v' \cdot p_\pi -\Delta \tildeM' }{v' \cdot q} + \frac{m_\pi^2-  (\Delta \tildeM')^2  - \Gamma^{\prime 2}/4}{2 M' v' \cdot q}  ~.
 \label{zstar-prime-gamma}
\end{align}
Just like in the narrow-width case, these relations can by used to derive,
\begin{align}  
    \frac{v\cdot k_\star'}{M'}  &= \frac{v\cdot p_\pi}{M'} - \frac{v'\cdot p_\pi - \Delta \tildeM'}{M}- \frac{ m_\pi^2- (\Delta \tildeM')^2- \Gamma^{\prime 2}/4}{2 M' M}~,\\
    \frac{v'\cdot k_\star}{M}& = \frac{v'\cdot p_\pi}{M} - \frac{v\cdot p_\pi + \Delta \tildeM}{M'}~+ \frac{ m_\pi^2- (\Delta \tildeM)^2- \Gamma^{2}/4}{2 M'M}.
\end{align}

At leading order the $z_{\star}^{(\prime)(0)}$ solutions are modified as
\begin{align}
    z_\star^{(0)} &= \frac{v \cdot p_\pi + \Delta M - \iu \, \Gamma/2}{v \cdot q} \,, \nn\\
    z^{\prime(0)}_\star &= \frac{v' \cdot p_\pi - \Delta M' + \iu \, \Gamma'/2}{v' \cdot q} ~.
\end{align}
Therefore, the $\hhchpt$ scalar propagators are now given by the tree-level result
\begin{equation}\label{eq:props-cms}
    2 D_v(-p_\pi) =\frac{\iu}{-v\cdot p_\pi - \Delta M + \iu \Gamma /2 }\,, \ \ \ 2 D_{v'}(p_\pi) =\frac{\iu}{v'\cdot p_\pi - \Delta M' + \iu \Gamma' /2 }\,.
\end{equation}

At $\order{1/M^{(\prime)}}$ the kinetic energy insertions originating from the double poles are also corrected by operators proportional to the finite widths. These modifications are described by the shifts
\begin{equation}
    \frac{\kpst^2}{2 M} \rightarrow \frac{\tilde{\vb{k}}_\star^2 - \Gamma^2/4}{2 M} \,, \ \ \ \
    \frac{\kpst^{\prime 2}}{2 M'} \rightarrow \frac{\tilde{\vb{k}}_\star^{\prime 2} - \Gamma^{\prime 2}/4}{2 M'} \,,  \label{eq:kinetic-gamma-shifts}
\end{equation}
where we have used the equations of motion $v^{(\prime)} \cdot k_\star^{(\prime)} - \Delta \tildeM^{(\prime)} =  \order{1/M^{(\prime)}}$ and $\tilde{\vb{k}}_\star^{(\prime) 2} = m_\pi^2 - (\Delta \tildeM^{(\prime)})^2$. 
Although in conventional schemes the finite width appears at 1-loop order, in the complex mass scheme the width is introduced already at the tree level. Therefore when employing the complex mass scheme it is consistent to simply shift the positions of the poles at $z_\star^{(\prime)(0)}$ and to ignore branch cuts. 

Strictly speaking the amplitudes on these resonance poles do not factorize anymore into a product of physical on-shell amplitudes as they have an external leg corresponding to an on-pole resonance, which is not part of the asymptotic Hilbert space. However, this is not a problem: a similar factorization still carries through as it has been shown in the proof of the unitarity of the complex mass scheme in relativistic theories\footnote{The proof relies on dressed propagators, loop corrected vertices and Veltman's largest time equation \cite{Veltman:1963th}. One particle irreducible graphs and subgraphs are computed in the zero-width limit. }~\cite{Denner:2014zga}.
Therefore, factorized expressions similar to \cref{eq:residue-1-noprime,eq:residue-1-prime} can still be written, with the modifications to $z_\star^{(\prime)}$ described above and with $\mathcal{A}_{L, R}$ now being Green's functions (with $n-1$ on-shell external particle legs and one ``on-pole'' unstable resonance leg).  In what follows we still refer to $\mathcal{A}_{L,R}$ as the left- and right-amplitudes, despite their strict interpretation as Green's functions rather than on-shell scattering amplitudes. 
Moreover, while the shift in the position of the poles shifts the momenta $k_\star^{(\prime)}$, as it has been shown in \cref{app:infinity}, the contribution from boundary term at infinity is equivalent to the replacement $k_\star^{(\prime)} \rightarrow k^{(\prime)}$ everywhere except in the kinetic energy insertions. Therefore, the only effect of the finite width on the BCFW factorization formula is given by \cref{eq:props-cms,eq:kinetic-gamma-shifts}.

We now move to discuss the modifications that the complex mass scheme introduces in the matching onto HQET.
For the weak current, $\mathcal{A}^{(h)}_{L, R}$, at leading order in $\Gamma/M$ the particle is stable and can therefore be matched directly onto the well-known HQET calculations performed treating the resonance as exactly stable \cite{Leibovich:1997em}. Since $\Gamma \ll \Lambda_\chi$, all the $\order{\Gamma/M}$ effects are fully captured within \hhchpt. 

The left- and right-amplitudes for the weak current are then given by \cref{vertex-correction-1,vertex-correction-2} but with a complex mass-splitting $\Delta M^{(\prime)} - \iu \Gamma^{(\prime)}/2$. Note that $M=M_{B}$ and $M'=M_{D}$ are still the real ground-state masses and therefore the normalization of the states is not affected by the complex mass deformation. 
Finally, deforming the hadron mass into the complex plane changes the matching of hadron mass parameters between \hhchpt and HQET.
In particular, for a hadronic multiplet ${\sf H}$, the definition of $\bar \Lambda_{\sf H}$, being the difference between the heavy hadron and quark masses at leading power, gets modified at next to leading power
\begin{equation}
   M_{{\sf H}} - m_Q = \bar \Lambda_{\sf H} + \order{1/m_Q} \ \ \rightarrow  \ \ \tildeM_{\sf H} - m_Q = \bar \Lambda_{\sf H} - \iu \Gamma_H/2+ \order{1/m_Q}. 
\end{equation}
and similar modification will be introduced to the mass parameters $\lambda_{1,2}$ at $\mathcal{O}(1/m_Q^2)$.
These mass parameters enter the weak matrix elements due to modified Schwinger-Dyson relations in HQET (or modified Ward identities, relating certain higher order IW functions to lower order ones, see e.g., Appendix B of Ref.~\cite{Bernlochner:2022ywh}).
They contain the mass differences between heavy hadrons and heavy quarks expanded in powers of $1/m_Q$. Therefore in the complex mass scheme, the mass difference between the heavy hadron and the heavy quark gets shifted by the complexification of the hadron state mass.
Its effects in the matching can be practically accounted for by taking the conventional HQET formula (e.g., Eqs.~(4.44) and (4.55) of Ref.~\cite{Manohar:2000dt}) and shifting in the Schwinger-Dyson terms~\cite{Bernlochner:2022ywh}, 
\begin{equation}
    ~\overline{\Lambda}_{\sf H}~\rightarrow ~\overline{\Lambda}_{\sf H}~ - \iu \Gamma_{\sf H}/2~. 
\end{equation}
To make finite-width effects explicit we can expand in $\Gamma^{(\prime)}$ (focusing e.g., on a $H_v \rightarrow S'_{v'}$ transition for the left-amplitude, and dropping the subscript ${\sf H}$) 
\begin{align}
    \begin{split}
    \label{vertex-correction-1-iGamma-exp}
    \mathcal{A}_{L}^{(h)}(0)&= \qty[\mathcal{A}_L^{(h)}(0)]_{\Gamma=0} -\qty(\frac{\iu \Gamma'}{2}    \pdv{\overline{\Lambda}'} + \frac{\iu \Gamma}{2}    \pdv{\overline{\Lambda}}) \qty[\mel{S'_{v'}(0)}{\hat{J}_\Gamma}{H_v(0)}_{1/m_{b,c}}]_{\Gamma=0} \\
    &\hspace{0.6\linewidth}+\order{1/M^2}~,
    \end{split}
\end{align}
The derivative with respect to $\overline{\Lambda}$ always produces a term proportional to the leading order Isgur-Wise function.\!\footnote{The derivative with respect to $\bar \Lambda^{(\prime)}$ only applies to the terms proportional to the leading order IW function as they are those originating from the Schwinger-Dyson relations. Any other occurrence of $\bar \Lambda^{(')}$, like those introduced to render the subleading IW functions dimensionless (e.g.\ $\chi_1 \rightarrow \overline{\Lambda} \chi_1$)~\cite{Bernlochner:2022ywh}, is spurious and should not be shifted.} 

In addition to the modifications discussed above, corrections appear at higher loop order from e.g., a shift in the self-energy, but this goes beyond tree level approximation in the complex mass scheme. We sketch how these corrections could be incorporated in \cref{NLO}.

\subsection{Example with off-shell $D^{(**)}$-mesons}
As a simple illustration, we consider the contribution of off-shell $D^{**}$ states to the amplitude $B\rightarrow D\pi\ell \nu$. This is interesting phenomenologically as some of the $D^{**}$ states have fairly large widths, and it also has the advantage that the analysis can be carried at $\mathcal{O}(1/M)$ (contrary to the case of the $D^*$, whose width is already higher order). We again focus only on the Born amplitude, leaving contact terms due to higher dimension operators to future work. For concreteness, we focus on the contribution from the $D_0^*$ resonance. 

Using the same analysis as in the preceding section, we may obtain the off-shell vertex correction for the $B\rightarrow D_0^*$ vertex. The Isgur-Wise function for the $D_0^*$ is denoted by $\zeta(w)$. Only the axial current contributes to the  $B-D_0^*$ vertex, 
\begin{equation}
    \begin{split}
    \label{D0star-example-weak}
    \mel{D_0^*(p_\pi)}{\hat{J}_A}{\bar{B}} &= \zeta(v\cdot v') \mel{D_0^*}{{\rm Tr}\big[~ \overline{S}'_{v'} \gamma^\nu\gamma_5 H_{v}~ \big]}{\bar{B}}_{1/m_{b,c}^0} \\
    &\hspace{0.1\linewidth}
    +\zeta'(v\cdot v') \frac{v\cdot p_\pi}{M'}\mel{D_0^*}{{\rm Tr}\big[~ \overline{S}'_{v'} \gamma^\nu\gamma_5 H_{v}~ \big]}{\bar{B}}_{1/m_{b,c}^0}  \\
    &\hspace{0.1\linewidth}
    +\zeta(v\cdot v') \frac{p_\pi^\mu}{M'} \pdv{v'_\mu} \mel{D_0^*}{{\rm Tr}\big[~ \overline{S}'_{v'} \gamma^\nu\gamma_5 H_{v}~ \big]}{\bar{B}}_{1/m_{b,c}^0} \\
    & \hspace{0.2\linewidth} 
    + \mel{D_0^*}{\hat{J}_A}{B}_{1/m_{b,c}^1} + \order{1/M^2}~. 
    \end{split}
\end{equation}
All of the finite width effects are contained in $\mel{D_0^*}{\hat{J}_A}{B}_{1/m_{b,c}^1}$. Using \cref{vertex-correction-1-iGamma-exp} one has, explicitly~\cite{Leibovich:1997em,Bernlochner:2017jxt}, 
\begin{equation}
    \begin{split}
    \mel{D_0^*}{\hat{J}_A}{B}_{1/m_{b,c}^1} = &\qty[ \mel{D_0^*}{\hat{J}_A}{B}_{1/m_{b,c}^1}]_{\Gamma=0}  \\[6pt]
    &~~~~~~-\bigg[\qty(\frac{\iu \Gamma'}{2}    \pdv{\overline{\Lambda}'} + \frac{\iu \Gamma}{2}    \pdv{\overline{\Lambda}})  \mel{D_0^*}{\hat{J}_A}{B}_{1/m_{b,c}^1}\bigg]_{\Gamma=0}\\[6pt]
    =&\qty[ \mel{D_0^*}{\hat{J}_A}{B}_{1/m_{b,c}^1}]_{\Gamma=0}  \\[6pt]
    &~~~~~~+\frac{\iu \Gamma'}{2}\frac{\zeta(w)}{1+w}\qty[\frac{3 w}{2 m_c} (v+v')^\nu+\frac{1+2w}{m_b}v^\nu]~,
    \end{split}
\end{equation}
where $w=v \cdot v'$ and we have used the $\Gamma = 0$ explicit matrix element at this order in the power expansion to compute the second line~\cite{Bernlochner:2017jxt}. 
The left-amplitude is related to the $D_0^* \rightarrow D \pi$ vertex, which must be evaluated to $\order{1/M}$, \emph{cf.} \cref{eq:Dstst-Dpi}.  

\subsection{Deviations from Breit-Wigner \label{NLO} }
The formalism we have presented so far has been formulated in the complex mass scheme at leading order in \hhchpt. For broad resonances, one may seek an improved description of the line shape beyond the Breit-Wigner approximation that is inherent to tree-level calculations in the complex mass scheme. More generally, one may be interested in how to compute loop corrections involving soft pions. Various approaches have been developed ranging from extending the complex mass scheme to loop-level calculations~\cite{Denner:2005fg} to developing effective theories for unstable particles~\cite{Beneke:2003xh,Beneke:2004km,Beneke:2015vfa}. 

The resonant states we consider are analogous to the $\rho$ meson in $\chi$PT, whose inclusion in the theory is theoretically complicated beyond tree level~\cite{Kampf:2009jh}. Nevertheless, resonance chiral theory (R$\chi$T) \cite{Ecker:1988te} is fairly successful in providing quantitative extrapolations up to energies $E\sim \Lambda_\chi$~\cite{Guerrero:1997ku,Pich:2001pj,GomezDumm:2003ku,Ruiz-Femenia:2003jdx,Cirigliano:2006hb}. Moreover, in our case, the linearity of the \hhchpt propagators and the heavy quark spin symmetry relating amplitudes of unstable hadrons with different spins within the same multiplet, considerably ameliorate these problems. In this Section, we sketch how a similar program could be pursued for the higher resonances such as the $D_0'$. 

The subject of one-loop corrections and their connection to on-shell amplitudes is a well-developed subject \cite{Brandhuber:2022qbk}. Using an exhaustive basis of one-loop master integrals \cite{Passarino:1978jh,tHooft:1978jhc}, one can reconstruct the coefficients of the mater integrals using a judicious choice of ``cuts''. This expresses the coefficients in the series in terms of on-shell lower order (i.e.~tree level) amplitudes~\cite{Bern:1996je,Elvang:2015rqa}. Such a program can be straightforwardly extended to the phenomenology we consider here: master integrals for one-loop functions are known~\cite{Bouzas:1999ug,Bouzas:2002xi,Zupan:2002je} and the treatment of on-shell tree-level amplitudes in \hhchpt and their factorization into lower point on-shell amplitudes has been the subject of this work.

While an exhaustive treatment of loop-level corrections is likely unnecessary, it may be of interest for characterizing the line shape of resonant states. In particular, deviations from a Breit-Wigner shape can be substantial and arise from iterated insertions of the resonant state's self-energy. Such a program has been successfully pursued in the context of R$\chi$T \cite{Guerrero:1997ku}. There, one makes use of dispersive arguments and a large-$N_c$ expansion. The leading-$N_c$ effect is fixed by the Weinberg sum rule and reproduces the classic vector meson dominance model. The next-to-leading $1/N_c$ correction can be obtained in the soft limit using chiral perturbation theory and can be extrapolated to larger values of $p$ using an Omn\`es resummation formula. Finally, making use of a calculation of the $\rho$ off-shell width $\Gamma_\rho(s)$, one can obtain an ansatz for the amplitude off-shell in which only the real part of the loop factor is included in the Omn\`es exponential and the imaginary part is shifted into the propagator. Other related approaches using unitarized $\chi$PT have also shown reasonable success phenomenologically \cite{Meissner:2020khl}. 

A similar analysis could be performed for resonant heavy meson excited states. It would be interesting to see if combining low energy input from \hhchpt alongside input from explicit resonance modeling could produce similar phenomenological successes. We leave this to future work.  

\section{Feynman rules description of the results}
\label{sec:Feynman}

We can summarize our findings via a set of prescriptive Feynman rules. One can readily check that these rules are self-consistent, i.e., that they satisfy \cref{eq:cauchy}. We draw heavy particle lines with a double line, pions with a single dashed line, kinetic energy insertions with a star, and the weak current with a square. The explicit Feynman rules for the ground-state multiplets are 
\begin{align}
    \label{feyn-propagator}
    \begin{split}
  \raisebox{5pt}{
    \begin{fmffile}{hhchipt-propagator}
    \begin{fmfgraph*}(75,10) 
    \fmfstraight
    \fmfbottom{i1,o1}
    \fmf{double,label=$\tilde{v}=v+k/M$}{i1,o1}
    \fmfv{label=${}_\alpha^{\dot\alpha}$}{i1}
    \fmfv{label=${}_\beta^{\dot\beta}$}{o1}
\end{fmfgraph*}
\end{fmffile} } 
  \quad &= \quad  
    -\frac{\iu}{v\cdot k-\delta \tildeM_H} \qty(\tilde \Pi_+)_\alpha^\beta \qty(\tilde \Pi_-)^{\dot{\alpha}}_{\dot{\beta}}  \\
    &\hspace{0.1\linewidth}+\iu\qty(\tilde \Pi_+ \gamma^5)_\alpha^{\dot{\alpha}} \qty(\gamma^5 \tilde \Pi_+)^{\beta}_{\dot{\beta}}\qty(\frac{1}{v\cdot k-\delta\tildeM_H}-\frac{1}{v \cdot k}) ~,
    \end{split}\\[10pt]
    \label{feyn-kin-insertion}
  \raisebox{5pt}{
    \begin{fmffile}{kinetic-insertion}
    \begin{fmfgraph*}(75,10) 
    \fmfstraight
    \fmfbottom{i1,v1,o1}
    \fmf{double}{i1,v1}
    \fmf{double}{v1,o1}
     \fmfv{d.sh=pentagram,d.si=3mm}{v1}
     \fmfv{label=${}_\alpha^{\dot\alpha}$}{i1}
    \fmfv{label=${}_\beta^{\dot\beta}$}{o1}
    \fmfv{label=$\tilde{v}=v+k/M$}{v1}
\end{fmfgraph*}
\end{fmffile} } 
   \quad &= \quad \iu \frac{k^2-(\Delta M)^2}{2 M} \delta_\alpha^\beta \delta_{\dot\beta}^{\dot\alpha}~, \\[10pt]
    \label{feyn-pion-vertex}
  \raisebox{5pt}{
    \begin{fmffile}{pion-vertex}
    \begin{fmfgraph*}(75,35) 
    \fmfstraight
    \fmfbottom{i1,v1,o1}
    \fmftop{ti,t1,to}
    \fmf{double,label=$\tilde{v},,p_\pi$,tension=1.75}{i1,v1}
    \fmf{dashes,label=$p_\pi$}{v1,t1}
    \fmf{double,label=$\tilde{v}$}{v1,o1}
    \fmf{phantom,tension=4}{i1,pvi}
    \fmf{phantom,tension=0.25}{pvi,ti}
    \fmf{phantom,tension=4}{o1,pvo}
    \fmf{phantom,tension=0.25}{pvo,to}
    \fmfv{label=${}_\alpha^{\dot\alpha}$}{pvi}
    \fmfv{label=${}_\beta^{\dot\beta}$}{pvo}    
\end{fmfgraph*}
\end{fmffile}}
\quad &= - \frac{g_H}{f} \tilde{v}\cdot\qty(\tilde{v}+\frac{p_\pi}{M})\delta_\alpha^\beta \,\qty[ \gamma^5\slashed{p}_\pi]_{\dot\beta}^{\dot \alpha}\quad ~,\\[24pt]
\raisebox{-5pt}{
    \begin{fmffile}{hard-current-onshell}
    \begin{fmfgraph*}(75,15) 
    \fmfstraight
    \fmfleft{i1}
    \fmfright{o1}
    \fmf{double,tension=1.75,label=$\tilde v$}{i1,v1}
    \fmf{double,label=$\tilde v'$}{v1,o1}
    \fmf{phantom}{v1,o1}
    \fmfv{d.sh=square,d.si=3mm}{v1}
    \fmfv{label=${}_\alpha^{\dot\alpha}$}{i1}
    \fmfv{label=${}_\beta^{\dot\beta}$}{o1}
\end{fmfgraph*}
\end{fmffile}}
\quad &= -\iu \xi(\tilde v \cdot \tilde v') \delta_{\dot \beta}^{\dot \alpha} \Gamma_\alpha^\beta +\iu\mel{H'_{v'}(0)}{\hat{J}_\Gamma(\tilde{v}',\tilde{v})}{H_{\tilde{v}}(0)}_{1/m_{b,c}} ~.\label{main-result-diagramtic}
\end{align}
Feynman rules with non-zero residual momenta (i.e., using $v,k$ instead of $\tilde{v}$) can be derived using RPI transformations. 
\cref{main-result-diagramtic} is a new result of this work. All the other Feynman rules have been obtained by extending the standard results in \hhchpt to their RPI-invariant form at $\mathcal{O}(1/M)$ (\cref{feyn-kin-insertion} is equivalent to the conventional $\vb{k}^2/2M$ by a field redefinition), and in the case of the propagator, including also the effects of the hyperfine mass (and width) splitting. Similar Feynman rules can be derived for the multiplets of the excited heavy hadron states.

The weak current Feynman rule \cref{main-result-diagramtic} could have been guessed based on RPI, while the constructive approach used in this paper offers a foundation for its validity. 
Our analysis can be summarized succinctly in terms of it, even though it stems from a non-trivial interplay between the proper reconstruction of factorization channels and a cancellation from a contour at large-$z$.

The BCFW-like construction presented herein relates the amplitude with an off-shell line, to two on-shell amplitudes. With both particles placed on-shell one can then use RPI to shift the finite residual momentum into a shifted reference vector $\tilde{v}'$. This procedure can be represented graphically as follows: Denote off-shell heavy particles by dashed lines, on-shell heavy particles by solid lines, and the weak current by a black square. Our analysis gives the Feynman rule for the half-off-shell vertex as 
\vspace{24pt}
\begin{equation}
    \\
  \raisebox{-15pt}{
    \begin{fmffile}{off-shell}
    \begin{fmfgraph*}(75,35) 
    \fmfstraight
    \fmfbottom{i1,o1}
    \fmftop{i2,t1,t2,t3,t4,o2}
    \fmf{double}{i1,t2}
    \fmf{dbl_dashes,label=$k$}{t2,o2}
    \fmfv{d.sh=square,d.si=3mm}{t2}
    \fmfv{label=$v$}{i1}
    \fmfv{label=$v'$}{o2}
\end{fmfgraph*}
\end{fmffile} } 
    =  
  \raisebox{-15pt}{
    \begin{fmffile}{on-shell}
    \begin{fmfgraph*}(75,35) 
    \fmfstraight
    \fmfbottom{i1,o1}
    \fmftop{i2,t1,t2,t3,t4,o2}
    \fmf{double}{i1,t2}
    \fmf{double,label=$k$}{t2,o2}
    \fmfv{d.sh=square,d.si=3mm}{t2}
    \fmfv{label=$v$}{i1}
    \fmfv{label=$v'$}{o2}
\end{fmfgraph*}
\end{fmffile} } 
    \quad\quad \rightarrow \quad\quad
    \raisebox{-15pt}{
    \begin{fmffile}{on-shell-rpi}
    \begin{fmfgraph*}(75,35) 
    \fmfstraight
    \fmfbottom{i1,o1}
    \fmftop{i2,t1,t2,t3,t4,o2}
    \fmf{double}{i1,t2}
    \fmf{double,label=$k=0$}{t2,o2}
    \fmfv{d.sh=square,d.si=3mm}{t2}
    \fmfv{label=$v$}{i1}
    \fmfv{label=$\tilde{v}'$}{o2}
\end{fmfgraph*}
\end{fmffile} } ~.
\end{equation}
\vspace{0pt}

\noindent The equality follows after properly including the subtraction of terms from $C_\gamma$. The arrow denotes an RPI transformation which shifts the $k$ dependence of the on-shell leg into a modified reference vector $\tilde{v}'$. This then enters into the Feynman rules via a shift in the Isgur-Wise function, $\xi(v\cdot \tilde{v}') \simeq \xi(v\cdot v') + \tfrac{v\cdot k}{M} \xi'(v\cdot v')$, which accounts for the shifted kinematics of the half off-shell vertex at $\mathcal{O}(1/M)$. 

Notice that $v \cdot \tilde v'$ ($\tilde v \cdot v'$) is nothing else than
\begin{equation}
 v \cdot \tilde v' \simeq \frac{p_B}{M_B} \cdot \frac{p_{D\pi}}{\sqrt{m_{D\pi}^2}}~, \quad \tilde v \cdot v' \simeq \frac{p_{B\pi}}{\sqrt{m_{B\pi}^2}} \cdot \frac{p_D}{M_D}~.
\end{equation}
with $p_{D\pi} = p_D + p_\pi$ ($p_{B\pi} = p_B - p_\pi$), up to pinched terms that are absorbed in higher dimensional operators. In other words, BCFW applied in the non-Lorentz invariant \hhchpt plus RPI (the remnant of Lorentz invariance in the low energy theory) enforce the form factors (IW functions) to be evaluated at the correct relativistic $q^2$ constructed with the off-shell momenta as one would have naively guessed.

\vfill
\pagebreak

\section{Summary and Outlook \label{Concl} }
In this paper, we have derived corrections that arise when a heavy particle is slightly off-shell due to the emission of soft final state particles (in our application pions). We show that these effects always enter at $\order{1/M}$, and include effects from both pion kinematics $\order{p_\pi/M}$, and finite width effects, $\order{\Gamma/M}$. They are fully determined by perturbative unitarity of \hhchpt, analyticity of \hhchpt amplitudes, and RPI. 

The corrections we have derived can be entirely expressed in terms of known Isgur-Wise functions and their derivatives without any need to introduce a complex reference velocity. This allows for their construction using HQET calculations performed with one-particle states performed in the zero-width limit.  For example in the case of $B\rightarrow D\pi\ell\nu$ via a $B^*$ or $D^*$ intermediate state, we find that 
these corrections give rise to shifts such as 
\begin{equation}
    \label{shift-in-conclusions}
    \xi(v\cdot v') \frac{1}{v'\cdot p_\pi} +  \frac{1}{M}\xi'(v\cdot v') \frac{v\cdot p_\pi}{v'\cdot p_\pi}~, \qq{and} \xi(v\cdot v') \frac{1}{(-v\cdot p_\pi)} +  \frac{1}{M'}\xi'(v\cdot v') \frac{v'\cdot p_\pi}{v\cdot p_\pi}~. 
\end{equation}
Our BCFW construction places the intuitive result that RPI fixes all such off-shell effects on firmer ground. An interesting technical byproduct of our work is the careful accounting for the contour at large-$z$, which is required to use modern on-shell methods in derivatively coupled effective theories such as \hhchpt. 

Notice that when $v\cdot p_\pi\rightarrow 0$ there is a pole for diagrams involving initial state emission, whereas when $v'\cdot p_\pi \rightarrow 0$ there is a pole for diagrams with final state emission. Since the Isgur-Wise function depends on $w=v\cdot v'$, a shift in $v' \rightarrow v' + k/M$ results in a shift $\xi(v\cdot v') \rightarrow \xi(v\cdot v') + \frac{v\cdot k}{M} \xi'(v\cdot v')$. This shift to the vertex is accompanied by an off-shell propagator $\iu/(v'\cdot k)$. At non-zero recoil, $v'\neq v$, the corrections in \cref{shift-in-conclusions} do not ``pinch''.  We therefore conclude that in a heavy particle theory with two reference vectors $v$ and $v'$ there exist unambiguous vertex corrections related to poles in the low energy theory. These unambiguous vertex corrections have a simple explanation in terms of the deformations of the argument $w$ of the IW functions (or equivalently of the form factors in the relativistic theory). Up to pinched terms they correspond to the (possibly off-shell) velocities of the incoming and outgoing legs.  

We, therefore, reach a more general conclusion about EFTs with multiple reference vectors. In these theories, certain off-shell vertex corrections may be associated with the residue of poles corresponding to physical factorization channels. We have presented a method to uniquely identify and fix these contributions, using the theory's analytic structure, RPI symmetry, and a BCFW-inspired momentum deformation. We have applied this method to the case of both stable asymptotic states, and unstable resonances. We conclude that factorizable vertex corrections can be obtained by a straightforward shift of the reference vector $v^{(\prime)}_\mu \rightarrow \tilde{v}^{(\prime)}_\mu$. This prescription is entirely fixed by RPI, unitarity, and the analytic structure of the low-energy theory.

The results of this paper are a necessary input for analyzing $B\rightarrow D \ell \nu (n\pi)$ at sub-leading power in HQET and their implementation in software tools utilized by the experiments to analyze their data. Recent literature has discussed the role of resonances in semi-leptonic $B$ decays with additional pions in the final state \cite{Gustafson:2023lrz,Manzari:2024nxr}. Some of our results overlap with Refs.~\cite{Gustafson:2023lrz,Manzari:2024nxr} while some are completely new. We now discuss similarities are differences between the results presented above and those derived in Refs.~\cite{Gustafson:2023lrz,Manzari:2024nxr}.

Reference \cite{Manzari:2024nxr} considers a similar BCFW analysis presented herein, but uses an alternative deformation ({\it cf}. \cref{app:alt-def}) that only allows access to resonance in the bottom-quark system. By combining deformations on the $c$ and $b$ systems separately one could identify all of the resonances in both the $c$ and $b$ systems. There are important conceptual and technical differences between Ref.~\cite{Manzari:2024nxr} and our work. On the technical level they work in QCD+HQET, whereas we work in HQET+\hhchpt, they do not consider off-shell effects for the bottom-quark system as they focus on the leading near-resonance contributions. Concerning the common contributions to the amplitude, the results of~\cite{Manzari:2024nxr} are the same as the ones derived here when evaluated at $z=z'_*$, up to higher order terms. This leads to the main conceptual difference between the two works. 

On the conceptual level \cite{Manzari:2024nxr} assumes, in the specific example analyzed in that work, that the ``pole at infinity'' gives a vanishing contribution and uses this requirement to determine the large-$w$ behavior of the IW function. While it is generically true that the non-spurious contributions to the boundary term can be made vanishing for sufficiently large $z$ extrapolations in the deformed momenta (i.e., into the deep-UV where perturbative QCD is applicable, which is BCFW constructible). In doing so for a realistic QCD spectrum, one picks up the full tower of QCD resonances as a series of (mostly unknown) poles at \emph{finite} $z$ which is presently not under control given the associated lack of data. Furthermore, certain factorization channels unreachable by a given momentum deformation will always manifest themselves as contributions to the pole-at-infinity\footnote{The above-mentioned missing factorization channel from the pion emission off the $B$ line provides an example of such effect. See \cref{app:alt-def} for more details.} and cannot be neglected. This is true even if one performs the calculation multiple times, with multiple deformations: in each individual calculation a spurious pole at infinity appears.  Therefore, the vanishing of the pole at infinity in full QCD invoked in~\cite{Manzari:2024nxr} can only provide a ``sum-rule'' constraint, requiring that a particular linear combination of IW functions associated with different resonance poles 
falls sufficiently fast at large-$w$. Whether it may be possible to sharpen this constraint and turn it into a constraint that can be imposed directly on each term of the sum by analyzing the positivity properties of such a linear combination has not been studied yet. As of now, the requirement that $\xi(w)$ falls as $1/w^2$ at large $w$ corresponds to modeling the hadronic form factor as a dipole. While it may be a reasonably justified model, it cannot be derived from first principles without additional implicit assumptions. Furthermore, given the finite range of experimentally accessible $w$ values, any other parameterization may be equally valid in fitting the data. 

One may better understand this issue in a related process, i.e., fully exclusive deeply virtual Compton scattering (DVCS)\footnote{The two processes are related in the sense that the weak current is replaced with an electromagnetic current, which is brought from the final to the initial state by crossing symmetry. Finite $\pln^2$ corresponds to finite virtuality of the incoming photon.}~\cite{Collins:1996fb}.
There, one can see that the large-$w$  behavior is the same as the one indicated in~\cite{Manzari:2024nxr}, but the amplitude in that kinematic regime does not involve the exchange of a single, lowest-lying hadronic state, which is the assumption of~\cite{Manzari:2024nxr}.
Conversely, in our approach, we are able to retain theoretical control by restricting the analysis within the regime of validity of \hhchpt and keeping the large-$z$ contour at finite radius. This allows us to explicitly include the large-$z$ contour without relying on any extrapolation nor implicit assumption beyond the domain of validity of \hhchpt. The unknown contributions from higher resonances neglected in~\cite{Manzari:2024nxr} are parameterized by higher-dimensional operators within the EFT that provide explicit contributions to the contour at infinity. The symmetries of the EFT restrict their explicit form and the order in the power expansion at which they enter. Furthermore, by working in \hhchpt, and using a deformation that accesses both the bottom- and charm-resonances, we find corrections that are: {\it i}) transparently related to derivatives of the Isgur-Wise function by RPI symmetry, and {\it ii)} involve both the bottom- and charm resonances (or equivalently both $\mathcal{A}_L^{(h)}$ and $\mathcal{A}_R^{(h)}$). 

Reference \cite{Gustafson:2023lrz} considers a completely general parameterization of $B\rightarrow D\pi \ell \nu$ using a partial wave expansion (specifically Eqs.~(1) and (5) of that reference). A set of model-independent constraints are then derived using unitarity bounds. In practice, to fit the data, a factorization ansatz (their Eqs.~(10)-(12)), expected to hold at leading order in the heavy-mass expansion, is used to separate the hadronic matrix elements into a form-factor that is convolved with a final-state interaction (FSI) model. The FSI model\footnote{When restricted to $p_\pi \ll \Lambda_\chi$ chiral perturbation theory is model-independent. For $p_\pi \sim \Lambda_\chi$, even when extended to be ``unitarized'', there are issues with its power counting, and consequently, any prescription is inherently model-dependent.} uses unitarized chiral perturbation theory \cite{Guo:2009ct}, a coupled channels analysis, and dispersion relations. The parameterization of the weak form factors is done in the (model-independent) BGL framework \cite{Boyd:1995sq}.

The effects addressed in this paper incorporate effects at sub-leading order in the heavy-mass expansion. They therefore cannot be captured by the factorization ansatz mentioned above, which holds only at leading power in $1/M$. Furthermore, being phrased in terms of \hhchpt, our results automatically implement HQS, can be systematically improved, and provide a framework for estimating theoretical uncertainties.

Our analysis has been restricted to subleading power, and we have worked at leading order in perturbation theory. Working at leading order ensures that the analytic structure of the theory contains only isolated poles. At higher orders branch cuts will appear. It would be interesting to understand how the analytic deformation presented here generalizes at higher orders in perturbation theory, but we defer such an analysis to future work. 

Future work will incorporate the off-shell corrections discussed here in a systematic evaluation of soft pion matrix elements within \hhchpt \cite{BDpilnu}. The methods outlined here may also be of interest for radiative semi-leptonic $B$ decays, and other heavy-heavy transitions with soft-particles in the final state.

\acknowledgments
We thank Andreas Helset, Aneesh Manohar, Clifford Cheung, Dean Robinson, and Mark Wise for insightful discussions. The work of RP \& MP is supported by the U.S. Department of Energy, Office of Science, Office of High Energy Physics, under Award No. DE-SC0011632 and by the Walter Burke Institute for Theoretical Physics. RP is supported by the Neutrino Theory Network under Award Number DEAC02-07CH11359.

\clearpage
\appendix

\section{Heavy hadron chiral perturbation theory 
\label{app:HHchiPT-defs}}

Heavy hadron chiral perturbation theory (\hhchpt) describes the interaction of heavy strongly interacting hadrons with (psuedo-)Goldstone bosons whose momentum satisfies $p\ll \Lambda_\chi$ where $\Lambda_\chi\sim 1~{\rm GeV}$ \cite{Burdman:1992gh,Yan:1992gz,Wise:1992hn,Cho:1992cf,Cho:1992gg,Falk:1992cx,Goity:1994xn,Boyd:1994pa,Boyd:1995pq}. The fields in the Lagrangian are direct interpolating operators for hadronic states (as opposed to microscopic quarks as in QCD or HQET). 

\subsection{States and fields}
States are labeled by their velocity $v$ and residual momentum $k$. For example, a $D$ meson with momentum $p_\mu=M_D v_\mu + k_\mu$ is written as $\ket{D_v(k)}$ and is normalized as \cite{Manohar:2000dt}
\begin{equation}
    \braket{D_{v'}(k')}{D_{v}(k)} = 2 v^0 \delta_{vv'} (2\pi)^3 \delta^{(3)}(\vb{k}-\vb{k}')~. 
\end{equation}
There is a factor of $\sqrt{m_H}$ when converting between relativistic normalization and the convention defined above i.e., $\ket{D(p)} = \sqrt{m_D} \ket{D_v(k)}$. 
Heavy particle fields are defined in terms of projectors $\Pi_+(v)= (1+\slashed{v})/2$. Multiplets related by HQS are assembled into superfields \cite{Falk:1991nq,Falk:1992cx,Casalbuoni:1996pg},
\begin{align}
    H=\Pi_+ h &= \Pi_+ \qty(\slashed{D}^*- \gamma_5 D)~,\\
    S=\Pi_+ s &= \Pi_+ \qty(\gamma_5\slashed{D}_1^{\prime} + D_0^*)~, \\
    T^\mu =\Pi_+ t^\mu &= \Pi_+ \qty(D_2^{*\mu\nu}\gamma_\nu - \sqrt{\tfrac32}D_1^\nu\gamma_5\qty[g^\mu_\nu-\tfrac13\gamma_\nu(\gamma^\mu-v^\mu)] ) ~.
\end{align}
We have specialized to the explicit states of the $D$ meson system that are of phenomenological interest. Similar formulae hold for the $B$ meson system. The superfields satisfy the following relations,
\begin{align}
\label{eq:vrels}
H = \slashed{v} H = - H \slashed{v}\,, \nn \\
S = \slashed{v} S =  S \slashed{v}\,, \nn \\
T^\mu = \slashed{v} T^\mu = - T^\mu \slashed{v}\,, \nn \\
v \cdot T =0 \,.
\end{align}

The Goldstone bosons arising from chiral symmetry breaking are parameterized by 
\begin{equation}
    \Sigma = \exp\qty( \frac{2 \iu M}{f})~,
\end{equation}
where $M$ is an $N\times N$ matrix for $SU(N)$ (for $SU(2)$ it is made up of $\pi^0$ on the diagonal and $\pi^\pm$ on the off-diagonal). The field $\Sigma$ transforms as $\Sigma \rightarrow L \Sigma R^\dagger$. It is convenient to introduce $\xi = \sqrt{\Sigma} = \exp\qty(\iu M/f)$  which transforms as $\xi \rightarrow L \xi U^\dagger = U \xi R^\dagger$ in terms of which one defines $\Vpi$ and $\Api$,
\begin{align}
    \Vpi_\mu &= \frac{\iu}{2}\qty(\xi^\dagger \partial_\mu \xi + \xi \partial_\mu \xi^\dagger)~,\\
    \Api_\mu &= \frac{\iu}{2}\qty(\xi^\dagger \partial_\mu \xi - \xi \partial_\mu \xi^\dagger)~.
\end{align}
The field $\Api_\mu$ transforms like the adjoint representation, $\Api_\mu \rightarrow U \Api_\mu U^\dagger$, while $\Vpi$ instead acts like a connection and can be used to define the covariant derivative, 
\begin{equation}
    D_\mu = \partial_\mu - \iu \Vpi_\mu~.    
\end{equation}
When acting on a field in the fundamental representation, $F$, we have 
\begin{equation}
    (D F)_a = \partial F_a -\iu \Vpi_{ab} F_b~, 
\end{equation}
whereas for a field in the anti-fundamental representation, $\bar{F}$, we have instead
\begin{equation}
    (D \bar{F})_a= \partial \bar{F}_a + \iu \bar{F}_b \Vpi_{ba}~. 
\end{equation}
%

\subsection{The \hhchpt Lagrangian}

The Lagrangian at leading power is given by \cite{Casalbuoni:1996pg, Falk:1992cx},
\begin{equation}\label{eq:lagr-LO}
    \begin{split}
    \mathcal{L}_{1/M^0} = \frac{f^2}{8}& {\rm Tr} \Big[(\partial \Sigma)^2\Big] 
    + {\rm Tr} \Big[H_v \Big(\iu v  \cdot D\Big)  \bar{H}_v   \Big]
     -{\rm Tr} \Big[S_v \Big(\iu v \cdot D +\Delta M_S - \iu \frac{\Gamma_S}{2}\Big) \bar{S}_v \Big] \\
    &- {\rm Tr} \Big[T_v^\mu\Big(\iu v \cdot D +\Delta M_T -\iu \frac{\Gamma_T}{2}\Big)  \bar{T}_{v,\mu}\Big]\\
    &+ g_H {\rm Tr} \Big[\bar{H}_v H_v \gamma_5\slashed{\Api} \Big]
    + g_S {\rm Tr} \Big[\bar{S}_v S_v \gamma_5\slashed{\Api} \Big]
    + g_T {\rm Tr}\Big[ \bar{T}_v^\mu T_{v,\mu}
    \gamma_5\slashed{\Api} \Big]~\\ 
    &+ g'_{ST} {\rm Tr}\Big[ \bar{S}_v T^\mu_v \gamma_5\Api_{\mu}  \Big] + g'_{HS} {\rm Tr} \Big[\bar{H} S \gamma_5 (v \cdot\Api) \Big] ~. 
    \end{split}
\end{equation}
We use a prime to denote an inter-multiplet transition and subscripts to indicate the participating heavy particle species.  The traces are performed both on the $SU(2)$ indices and Dirac indices whenever appropriate. The $g'_{HT}$ coupling is absent at leading power \cite{Falk:1992cx}. At sub-leading power, the Lagrangian contains, among other terms, a kinetic energy operator,
\begin{equation}
    \mathcal{L}_{1/M} \supset  -\frac{1}{2M} {\rm Tr}\Big[ \bar{H}_v  D_\perp^2 H_v \Big] - \frac{1}{2M} {\rm Tr}\Big[ \bar{S}_v  (D_\perp^2 - \Delta M_S \Gamma_S) S_v\Big]  - \frac{1}{2M} {\rm Tr} \Big[\bar{T}_v^\mu (D_\perp^2 - \Delta M_T \Gamma_T)T_{v ,\mu}\Big]~.
\end{equation}
This can be related to a naive expansion of $D^2=(v\cdot D)^2+D_\perp^2$ via a field redefinition, i.e., using the equations of motion. For the excited multiplets, $\Delta M_X$ denotes the spin-averaged residual mass.
To fix our notation, at subleading power, we also list the heavy quark spin symmetry violation mass operator inducing the hyperfine mass splitting within a \hhchpt multiplet~\cite{Falk:1995th}, which we indicate with $\delta M_X$ ($ \sim {\cal O}(1/M)$)
\begin{equation}
    \mathcal{L}_{1/M} \supset  -\frac{\delta M_H}{8} {\rm Tr}\Big[ \bar{H}_v  \sigma^{\mu\nu} H_v  \sigma_{\mu\nu} - 6  \bar{H}_v H_v\Big] + \frac{\delta M_S}{8} {\rm Tr}\Big[ \bar{S}_v  \sigma^{\mu\nu} S_v \sigma_{\mu\nu}\Big]  + \frac{3\delta M_T}{16} {\rm Tr} \Big[\bar{T}_v^\alpha \sigma^{\mu\nu}T_{v ,\alpha} \sigma_{\mu\nu}\Big]~.\label{eq:lagr-hyperfine}
\end{equation}
where we have included the effect of a finite wave function renormalization for the ground-state multiplet to keep the $B_v$ and $D_{v'}$ mesons massless in the effective theory~\cite{Falk:1995th} and neglected a possible $S$-$T^\mu$ mass mixing term. As is well known, upon matching onto HQET, the values of these hyperfine mass splittings can be expressed in terms of the parameter $\lambda_2$. In the complex mass scheme the mass is shifted by $\delta M_X \rightarrow \delta \tildeM_X = \delta M_X - \iu \delta \Gamma_X/2$, which encodes a relative change in the decay widths of the mesons within a multiplet.

\subsection{Sums over states  and propagators \label{sec:pol-sum}}
Summing over intermediate particles and polarizations in \cref{Born-example-v1} reproduces the relevant propagators in the low energy effective theory. The polarization states combine with the denominator $\iu/(2 z_\star^{(\prime)} v^{(\prime)}\cdot q)$ and reproduce the spectral representation of the propagator. Using \cref{vertex-correction-1,vertex-correction-2}, which are explicitly expanded about zero residual momentum, all resulting propagators agree with \hhchpt Feynman rules \cite{Cho:1992cf,Manohar:2000dt}. We defer a discussion of the interplay between propagators and vertices under an RPI transformation to \cref{app:RPI-propagator}.

When considering a mass-degenerate multiplet, propagators can be assembled into a ``single trace'' contribution \cite{Cho:1992cf}. 
For hypermultiplets $H$, $S$ used in this paper with velocities $v$ and residual momentum $k$ flowing in the propagator one has
\begin{align}
  \sum_{{\rm states}\, i \in H} \sum_{s \in {\rm spins}(i)}  \frac{\mel{0}{H_{ v,\alpha}^{\dot{\alpha}}}{H^{i,s}_{ v}(0)}\mel{H^{i,s}_{ v}(0)}{~\overline{\!H}^{\beta}_{ v,\dot{\beta}}}{0}}{2 v\cdot k} &= -\frac{\iu \, \qty( \Pi_+)_\alpha^\beta \qty( \Pi_-)^{\dot{\alpha}}_{\dot{\beta}}}{v\cdot k}  + \mathcal{O}(1/M), \\
  \sum_{{\rm states}\, i \in S} \sum_{s \in {\rm spins}(i)}  \frac{\mel{0}{S_{ v,\alpha}^{\dot{\alpha}}}{S^{i,s}_{ v}(0)}\mel{S^{i,s}_{ v}(0)}{~\overline{\!S}^{\beta}_{ v,\dot{\beta}}}{0}}{2 (v\cdot k - \Delta \tildeM_S)} &= -\frac{\iu\, \qty( \Pi_+)_\alpha^\beta \qty( \Pi_+)^{\dot{\alpha}}_{\dot{\beta}}}{v\cdot k -\Delta \tildeM_S}  + \mathcal{O}(1/M)~,
\end{align}
where $\Delta \tildeM_i \equiv \Delta M_i -\iu \Gamma_i/2$ and we have suppressed light flavor indices.  We have made explicit the Dirac indices on the fields $H^{\dot \alpha}_\alpha$, $\bar H^{\beta}_{\dot \beta}$ (and similarly for $S$), as they appear in the traces of the left and right amplitudes. The numerators combine the left and right amplitudes in a single super-trace, joined by velocity projectors ({\it cf.}~the Feynman rules in Ref.~\cite{Cho:1992cf} for the $H$ propagator) and eliminate the need for defining deformed polarization vectors for the intermediate states. %

At $\mathcal{O}(1/M)$ one needs to consider the presence of a finite mass splitting. It is convenient to decompose the propagator as the sum of a ``single trace'' and ``double trace'' contributions.\!\footnote{In principle at this order there is also a mass mixing term between the spin-1 components of $S_v$ and $T_v^\mu$, \cref{eq:lagr-hyperfine}.  Its inclusion is straightforward but renders the expressions more complicated as one needs to consider a $2\times 2$ propagator matrix for the two multiplets simultaneously. Therefore we will neglect it in this discussion.} We will also consider a modified velocity $\tilde v$ in the matrix elements in the numerators to allow for their residual momentum dependence at $\mathcal{O}(1/M)$, which will be discussed further in \cref{app:RPI-propagator},

\begin{align}
  \sum_{{\rm states}\, i \in H} \sum_{s \in {\rm spins}(i)}  \frac{\mel{0}{H_{ \tilde v,\alpha}^{\dot{\alpha}}}{H^{i,s}_{\tilde v}(0)}\mel{H^{i,s}_{\tilde v}(0)}{~\overline{\!H}^{\beta}_{\tilde v,\dot{\beta}}}{0}}{2 (v\cdot k - \delta \tildeM_i)} &= 
  \begin{multlined}[t]
  -\frac{\iu \, \qty( \tilde \Pi_+)_\alpha^\beta \qty(\tilde  \Pi_-)^{\dot{\alpha}}_{\dot{\beta}}}{v\cdot k - \delta \tildeM_H}\\
   \hspace{-0.2\linewidth}+\iu \, \qty( \Pi_+ \gamma^5)_\alpha^{\dot \alpha} \qty( \gamma^5 \Pi_+)^{\beta}_{\dot{\beta}}\Big(\frac{1}{v\cdot k - \delta \tildeM_H} - \frac{1}{v\cdot k}\Big)\\ 
   +  \mathcal{O}(1/M^2)~,
   \end{multlined} \label{superfield-propagator}\\
  \sum_{{\rm states}\, i \in S} \sum_{s \in {\rm spins}(i)}  \frac{\mel{0}{S_{ \tilde v,\alpha}^{\dot{\alpha}}}{S^{i,s}_{\tilde v}(0)}\mel{S^{i,s}_{\tilde v}(0)}{~\overline{\!S}^{\beta}_{\tilde v,\dot{\beta}}}{0}}{2 (v\cdot k - \Delta \tildeM_S  -\delta \tildeM_i)} &= 
\begin{multlined}[t]
   \frac{\iu\, \qty(\tilde \Pi_+)_\alpha^\beta \qty( \tilde \Pi_+)^{\dot{\alpha}}_{\dot{\beta}}}{v\cdot k -\Delta \tildeM_S -\delta \tildeM_S/4 } \\
   \hspace{-0.4\linewidth}-\iu \, \qty( \Pi_+ )_\alpha^{\dot \alpha} \qty(  \Pi_+)^{\beta}_{\dot{\beta}}\Big(\frac{1}{v\cdot k - \Delta \tildeM_S  -\delta \tildeM_S/4} - \frac{1}{v\cdot k-\Delta \tildeM_S + 3 \delta \tildeM_S /4}\Big)\\ 
   + \mathcal{O}(1/M^2)~,  
\end{multlined}
\end{align}
where $\tilde \Pi_\pm = (1 \pm \slashed{\tilde v})/2$, $\delta \tildeM_i \equiv \delta M_i -\iu \,\delta \Gamma_i/2$. In deriving these formulae we just added and subtracted the matrix elements of one of the states divided by the propagator denominator of the other, choosing the ones having the simplest matrix elements (i.e. the lowest spin). The second terms, which are all ``double trace'' contributions, are all $\mathcal{O}(1/M)$ since they encode the effects of the hyperfine mass (and width) splitting. 
We have kept them written in terms of resummed propagators because they can better describe the lineshapes in decays, which is  phenomenologically desirable. If one wishes to expand them and keep the leading $1/M$ correction, it is easy to show that they assume the form of an insertion of the ${\cal O}(1/M)$ HQET Lagrangian \cref{eq:lagr-hyperfine} multiplied by two leading order propagators. Note, that in the specific example of $B\to D \ell \nu \pi$ the double trace contribution vanishes as there is no  $D-D-\pi$ ($B-B-\pi$) vertex in \hhchpt. 

\section{Reparameterization invariance }\label{app:RPI}

Under a reparameterization by 
\begin{align}
\label{eq:RPI-transform}
    k & \rightarrow  k-\epsilon \nn \\
    v & \rightarrow  v + \epsilon / M~,
\end{align}
the theory must remain invariant \cite{Luke:1992cs}. The 4-vector $q^\mu$ is further constrained by  $q \cdot v = q^2/2M$ to maintain the unit normalization of the velocity. This is a by-product of hidden Lorentz invariance and has non-trivial consequences for the effective Lagrangian. We briefly review the construction of building blocks that respect RPI and provide the relevant objects at $\order{1/M}$. 

In the main text, we perform RPI transformations to obtain matrix elements with external states with vanishin residual momentum. For the ground state multiplet, choosing $\epsilon = k$, one can set the residual momentum of the off-shell state to zero at the price of shifting its corresponding velocity. For an excited state with a finite residual mass $\Delta M$ one can choose $\epsilon$ such that
\begin{align}
  & k \rightarrow \Delta M \tilde v\,, \nn \\
  & v \rightarrow \tilde v\,, \\
  & \tilde v \cdot \tilde v = 1~,\nonumber
\end{align}
which can be solved for $\epsilon$:
\begin{equation}
    \epsilon_\mu= (k_\mu - \Delta M v_\mu)\frac {M}{M+\Delta M} \simeq k_\mu - \Delta M v_\mu +\order{1/M} ~.
\end{equation}
In the shifted basis we have $p_\mu  = (M + \Delta M ) \tilde v_\mu$ such that the residual momentum is necessary to reconstruct the full excited hadron mass. Matrix elements can therefore be directly matched to HQET calculations where the heavy quark velocities $v^\mu$ are chosen conventionally as $p_H^\mu = M_H v^\mu$.

Note also that performing an RPI transformation on one leg of the vertex but not the other results in a residual phase $\e^{\iu k_\star \cdot x}$ multiplying the hadronic matrix element. However, as made apparent by keeping the Fourier transform in \cref{eq:J-weak}, this phase is removed by exact momentum conservation with the leptonic system.

\subsection{RPI building blocks \label{app:RPI-BB}}

Here we define fields and labels that only acquire a phase under an RPI transformation. These fields can be used to construct matrix elements that are manifestly invariant under RPI transformations.

Under an RPI transformation, the relevant objects transform as follows 
\begin{align}
    {\rm reference~ vector:}~~ v &\rightarrow v-q/M~,\\
    {\rm (pseudo)scalar~ field:}~~    P_v &\rightarrow \e^{-\iu q\cdot x} P_v~,\\
    {\rm (pseudo)vector~ field:}~~    A^\mu_v &\rightarrow \e^{-\iu q\cdot x} 
    \qty(A^\mu_v + \frac{1}{M} v^\mu q_\sigma A^\sigma_v),\\
    {\rm tensor~ field:}~~   T^{\mu\nu}_v &\rightarrow \e^{-\iu q\cdot x}  \qty(T^{\mu\nu}_v + \frac{1}{M} v^\mu q_\sigma T^{\sigma \nu}_v+ \frac{1}{M} v^\nu q_\sigma T^{\mu \sigma}_v) .
\end{align}
The building blocks that respect RPI at $\order{1/M}$ are given by~\cite{Cho:1993su,Luke:1992cs,Boyd:1994pa,Falk:1995th}
\begin{align}
    \mathcal{H}&= H + \frac{\iu }{2M} \qty[ \slashed{D},H] + \order{1/M^2_H}~,\\
    \mathcal{S}&= S + \frac{\iu}{2M} \qty{\slashed{D},S} + \order{1/M^2_S}~,\\
    \mathcal{T}^\mu & = T^\mu + \frac{\iu }{2M} \qty[ \slashed{D},T^\mu]-v^\mu \frac{\iu D \cdot T}{M_T}+ \order{1/M^2_T}~. 
\end{align}
These objects are RPI up to terms of $\order{1/M^2}$.

In \cref{NWA} we are interested in the case where the residual momentum is proportional to the velocity label $v^\mu$. In this case, one can use the relations \cref{eq:vrels} one finds
\begin{align}
\bra{0}\mathcal{H}_{v}(x)\ket{H_v(\Delta M_H v)} &\simeq e^{-\iu \Delta M_H v \cdot x}\frac{M_H}{ M} \bra{0}H_v(x)\ket{H_v(0)} + \order{1/M^2}\,, \\
\bra{0}\mathcal{S}_{v}(x)\ket{S_v(\Delta M_S v)} &\simeq e^{-\iu \Delta M_S v \cdot x} \frac{M_S}{ M} \bra{0}S_v(x)\ket{S_v(0)} + \order{1/M^2}\,, \\
\bra{0}\mathcal{T}^\mu_{v}(x)\ket{T_v(\Delta M_T v)} &\simeq e^{-\iu \Delta M_T v \cdot x} \frac{M_T}{M} \bra{0}T^\mu_v(x)\ket{T_v(0)} + \order{1/M^2}\,,
\end{align}
where $M_{H,S,T}=M + \Delta M_{H,S,T}$ and we have used the notation $\ket{H_v,S_v,T_v (k)}$ to indicate a state in the multiplet of $H$, $S$, $T$ with velocity $v$ and residual momentum $k$. The factors $M_{H,S,T}/M$ account for the different wave-function renormalizations and state normalizations i.e., $\frac{M+\Delta M}{M} = \sqrt{\frac{M+\Delta M}{M}} \times \sqrt{\frac{M+\Delta M}{M}}$, in a theory with a residual mass, where only $M v^\mu$ modes have been integrated out and a theory with no residual masses, i.e. where $M_{H,S,T} v^\mu$ modes have been integrated out and are necessary to correctly match \hhchpt to HQET.

\subsection{Propagators and vertices \label{app:RPI-propagator} }
Having introduced RPI building blocks, we can construct propagators using these fields and generalize \cref{superfield-propagator}. This effectively allows us to perform RPI transformations on the vertices connected to the internal propagator in a diagrammatic language. 

The strategy pursued in the main text involved performing RPI transformations on both the left- and right-amplitudes. This left us with a sum over on-shell states with zero residual momentum, and a shifted reference vector $\tilde{v}^{(\prime)}$. This leads to the analog of \cref{superfield-propagator} but with shifted reference vectors in the numerator, 
\begin{equation}
    \label{superfield-propagator-vtilde}
    \begin{split}
  \sum_{i \in (B,B^*)} &\sum_{s \in {\rm spins}(i)}  \frac{\mel{0}{{\cal H}_{\tilde v_i,\alpha}^{\dot{\alpha}}}{H^{i,s}_{\tilde v_i}(0)}\mel{H^{i,s}_{\tilde v_i}(0)}{~\overline{\! \cal H}^{\beta}_{\tilde v_i,\dot{\beta}}}{0}}{z_{\star,i} v\cdot q}  \\
  &= -\frac{\iu}{v\cdot p_\pi+\Delta M} \qty(\tilde \Pi_+)_\alpha^\beta \qty(\tilde \Pi_-)^{\dot{\alpha}}_{\dot{\beta}}  -\qty(\tilde \Pi_+ \gamma^5)_\alpha^{\dot{\alpha}} \qty(\gamma^5 \tilde \Pi_+)^{\beta}_{\dot{\beta}}\frac{\iu \Delta M}{(v\cdot p_\pi)(v \cdot p_\pi +\Delta M)}\\
  &\hspace{0.5\linewidth}+ \order{1/M^2}~,
  \end{split}
\end{equation}
where we have use the property ${\cal H}_{\tilde v}\ket{H_{\tilde v}(0)} = H_{\tilde v}\ket{H_{\tilde v}(0)}$ and defined the modified projectors ${\tilde \Pi}_\pm = (1 \pm \slashed{\tilde v})/2$. \Cref{superfield-propagator-vtilde} is used to derive \cref{explicit-sum-states-1,explicit-sum-states-2} of the main text. 

Alternatively, one can choose to not perform a RPI transformation, and instead retain the same residual momentum on the both states. The construction of the RPI building blocks guarantees that  
\begin{equation}
    \label{superfield-propagator-kres}
    \begin{split}
    \mel{0}{{\cal H}_{ v_i,\alpha}^{\dot{\alpha}}}{H^{i,s}_{ v_i}(k)}\mel{H^{i,s}_{ v_i}(k)}{~\overline{\! \cal H}^{\beta}_{ v_i,\dot{\beta}}}{0} =  \mel{0}{{\cal H}_{\tilde v_i,\alpha}^{\dot{\alpha}}}{H^{i,s}_{\tilde v_i}(0)}\mel{H^{i,s}_{\tilde v_i}(0)}{~\overline{\! \cal H}^{\beta}_{\tilde v_i,\dot{\beta}}}{0}  ~.
  \end{split}
\end{equation}
Both formulations therefore lead to equivalent results.

\section{Soft pion matrix elements \label{app:pion-mels} }
The pion matrix element for transitions in the ground state multiplet is 
\begin{equation}
    \begin{split}
    \mel{\pi(k) H_v(0)}{ {\rm Tr} \qty[\mathcal{H}_v \gamma_5 \slashed{\Api} \overline{\mathcal{H}}_v]}{H_v(k)}  = & g_H \mel{\pi(k) H_v(0)}{ {\rm Tr} \qty[H_v \gamma_5 \slashed{\Api} \overline{H}_v]}{H_v(k)}\\
    &+ \frac{g_H}{2M} \mel{\pi(k) H_v(0)}{ {\rm Tr} \qty[ [\iu \slashed{D},H_v] \gamma_5 \slashed{\Api} \overline{H}_v]}{H_v(k)} ~.
    \end{split}
\end{equation}
Then, using the identities $ \Pi_+\gamma^\mu \Pi_+ = -\Pi_- \gamma^\mu \Pi_- = v^\mu$, $\Pi_+ H_v=H_v$, and $H_v\Pi_-=H_v$ one can show that 
\begin{equation}
\Pi_+ [\gamma^\mu, H_v] \Pi_- = 2 v^\mu H_v~.
\end{equation}
Using this relation we have 
\begin{align}
    \mel{\pi(k) H_v(0)}{ {\rm Tr} \qty[\mathcal{H}_v \gamma_5 \slashed{\Api} \overline{\mathcal{H}}_v]}{H_v(k)}  
    =  & g_H\mel{\pi(k) H_v(0)}{ {\rm Tr} \qty[H_v \gamma_5 \slashed{\Api} \overline{H}_v]}{H_v(k)}\\
    &+ \frac{g_H}{M}\mel{\pi(k) H_v(0)}{ {\rm Tr} \qty[(\iu v \cdot D H_v) \gamma_5 \slashed{\Api} \overline{H}_v]}{H_v(k)}\nonumber\\
    =  & g_H \qty[1+ \frac{v \cdot k}{M}]\mel{\pi(k) H_v(0)}{ {\rm Tr} \qty[H_v \gamma_5 \slashed{\Api} \overline{H}_v]}{H_v(k)}~.\nonumber
\end{align}

Notice that the $1/M$ correction pinches a propagator (up to terms of $\mathcal{O}(1/M^2)$ and it effectively induces a contribution to a contact term. If we consider a resonance with mass $\Delta M$ decaying to a pion and a ground state heavy meson then the $v\cdot k$ terms in the numerator will not pinch the associated propagator. As an explicit example, we take the relevant multiplets for the $D_0^* \rightarrow D \pi$ matrix element, 
\begin{align}
    &\mel{\pi(k) H_v(0)}{ {\rm Tr} \qty[\mathcal{S}_v \gamma_5 (v \cdot\Api) \overline{\mathcal{H}}_v]}{S_v(k)}  =  g_{HS}' \mel{\pi(k) H_v(0)}{ {\rm Tr} \qty[ S_v \gamma_5 (v \cdot\Api) \overline{H}_v]}{S_v(k)}\\
    &\hspace{0.375\linewidth}+\frac{g_{HS}'}{2M} \mel{\pi(k) H_v(0)}{ {\rm Tr} \qty[ \{\gamma^\mu ,\iu D_\mu S_v \} \gamma_5 (v \cdot\Api) \overline{H}_v]}{S_v(k)}~.\nonumber
\end{align}%
Next using the identities $ \Pi_+\gamma^\mu \Pi_+ =  v^\mu$, $\Pi_+ H_v=H_v$, and $H_v\Pi_-=H_v$, $\Pi_+ S_v=S_v$ and $S_v\Pi_+=S_v$ we get
\begin{equation}
\Pi_+ \{\gamma^\mu, S_v\} \Pi_+ = 2 v^\mu S_v~.
\end{equation}
leading to 
\begin{equation}
    \begin{split}
    \mel{\pi(k) H_v(0)}{ {\rm Tr} \qty[\mathcal{S}_v \gamma_5 (v \cdot\Api) \overline{\mathcal{H}}_v]}{S_v(k)}      
    =  g_{HS}' \qty[1+\frac{v\cdot k}{M}] &\mel{\pi(k) H_v(0)}{ {\rm Tr} \qty[ S_v \gamma_5 (v \cdot\Api) \overline{H}_v]}{S_v(k)}  ~. \label{eq:Dstst-Dpi}
    \end{split}
\end{equation}
Notice that now the $v \cdot k/M$ does not pinch against the $\iu/(v \cdot k -\Delta \tildeM)$ propagator due to the presence of the residual mass.

\section{Explicit solution for deformed momentum \label{app:explicit-soln}} 
To find a solution for $q_\mu$ which satisfies $q\cdot p_{\ell \nu}=0$, $q\cdot p_\pi=0$, and $q^2=0$ it is convenient to 
define the auxiliary 4-vector $u_\mu$:
\begin{equation}
u = ( 0, \, \vec{p}_{\ell\nu} \cross \vec{p}_{\pi})~,
\end{equation}
which trivially satisfies the condition $u \cdot p_\pi = u \cdot \pln = 0$.
Then we can write 
\begin{equation}
    \label{expl-q-soln}
    q_\mu = \mathfrak{z}\big(\iu\, \epsilon_{\mu\nu\rho\sigma}\,\pln^\nu \, p_\pi^\rho\, u^\sigma \pm \sqrt{(\pln \cdot p_\pi)^2 - m_{\ell\nu}^2 \, m_\pi^2} \, u_\mu \big)~,
\end{equation}
where $m_{\ell\nu}^2=(p_\ell+p_\nu)^2$ is the invariant mass of the neutrino-lepton system. The expression above provides two solutions up to a rescaling by some complex number $\mathfrak{z}$. The conditions $q\cdot p_{\ell \nu}=0$ and $q\cdot p_\pi=0$ are satisfied by construction, while the condition $q^2=0$ fixes the coefficient of the second term as can be verified with a bit of algebra.

The sign in \cref{expl-q-soln} is equivalent to a convention for the labeling of spin states (i.e.\ spin up vs. down) along an axis perpendicular to the plane spanned by the $\ell\nu$ and $\pi$ momenta, and this convention does not affect physical results. We note that the combinations $v\cdot k_\star$ or $v'\cdot k_\star$ are entirely dictated by external kinematics. 

\subsection{Uniqueness of solution \label{app:uniqueness}}

Despite appearances, $k_\star$ is in fact independent of $\mathfrak{z}$. To see this notice that only the ratio $q/v'\cdot q$ or $q/v\cdot q$ appears in the definition of $k_\star$, 
\begin{equation}
    k^{(\prime)}_\star = p_\pi - z^{(\prime)}_\star q = p_\pi - \frac{v^{(\prime)}\cdot p_\pi \pm\Delta M^{(\prime)} }{v^{(\prime)}\cdot q} q~.
\end{equation}
Therefore, under a $c$-number rescaling $q\rightarrow \mathfrak{z} q$ it follows that $k_\star$ is fixed independent of the choice made for $\mathfrak{z}$. The off-shell matrix elements obtained via the BCFW-like reduction formula are then also independent of $\mathfrak{z}$.

\subsection{Construction of polarization vectors \label{app:pol-vecs}}
In this section, we show that any contraction of $k_\star$ with a polarization vector can be rewritten in terms of $v\cdot k_\star$ and $v'\cdot k_\star$. While not directly used in the examples presented in this paper, this can be useful for amplitudes where the heavy hadron in the initial or final state has a spin greater than zero. We will focus on the spin-1 case but similar arguments can be applied to different values of the particle's spin. Let us, without loss of generality, consider the emission of a pion from the leg labeled by $v$. A complete, but not orthonormal, basis that spans the space perpendicular to $v$ is given by $\{ \vb{v}' , \vb{p}_{\ell\nu}, \vb{p}_\pi\}$ where the boldface denotes the space-like vector orthogonal to the time-like direction $v$. 

Any contraction of $k_\star$ with a polarization vector $\epsilon^{(i)}$ can be written, for three complex numbers $\alpha_1$, $\alpha_2$, and $\alpha_3$ as
\begin{equation}
    \begin{split}
    k_\star \cdot \epsilon &=~ \alpha_1 \kpst \cdot \vb{v}' + \alpha_2 \kpst \cdot \vb{p}_{\ell\nu} + \alpha_3 \kpst \cdot \vb{p}_{\pi} \\
    &=~(\alpha_1+\alpha_2+\alpha_3) v\cdot k_\star -\alpha_1 v'\cdot k_\star   -\alpha_2 \pln\cdot  k_\star -\alpha_3 p_{\pi}\cdot k_\star~. 
    \end{split}
\end{equation}
Next using $q\cdot \pln=0$ and $q\cdot p_\pi=0$ we have 
\begin{equation}
    k_\star \cdot \epsilon =(\alpha_1+\alpha_2+\alpha_3) v\cdot k_\star -\alpha_1 v'\cdot k_\star   -\alpha_2 \pln\cdot  p_\pi -\alpha_3 p_{\pi}^2~. 
\end{equation}
The coefficients $\alpha_i$ depend on the kinematic variables $v\cdot v'$, $v\cdot p_\pi$, and $v'\cdot p_\pi$. 

\subsection{Lepton currents with no $z$-dependence \label{app:lnu-deform}}

Here we show that it is safe to ignore the leptonic current in the analysis presented in this work as it does not introduce any additional dependence on the complex deformation parameter $z$. Namely, for each choice of a Dirac structure $\Gamma'$, we can choose how to share the deformation $\pln^\mu \rightarrow \pln^\mu +z q^\mu$ between the lepton and neutrino momenta $p_\ell^\mu$, $p_\nu^\mu$ such that the matrix element of the leptonic current $J_{\Gamma'}$ between the charged lepton and (anti-)neutrino does not depend on $z$. 
Namely, we seek a deformation
\begin{equation}
 p_\ell^\mu \rightarrow p_\ell^\mu + z q_\ell^\mu~, \quad p_\nu^\mu \rightarrow p_\nu^\mu + z q_\nu^\mu~,
\end{equation}
with $q_{\ell, \nu}$ satisfying
\begin{equation}
 q_{\ell, \nu}^2 =0, \quad p_\ell \cdot q_\ell =0, \quad p_\nu \cdot q_\nu = 0, \quad q_\ell^\mu + q_\nu^\mu = q^\mu~.
\end{equation}
 The first three conditions guarantee that the charged lepton and neutrino are kept on-shell, the last are required for compatibility with \cref{eq:deform-conditions}. 
Since the neutrino is massless and left-handed, the leptonic matrix element, when expressed in the (massive) spinor helicity formalism, will have one of the two forms
\begin{equation}\label{eq:lept-me-forms}
 [{\bf \ell}^{\bf L}| J_{\Gamma'} | \nu \rangle~, \qquad \langle{\bf \ell}^{\bf L}| J_{\Gamma'} | \nu \rangle~,
\end{equation}
where we have used the massive spinor helicity notation of~\cite{Arkani-Hamed:2017jhn}.
The first matrix element corresponds to $\Gamma' = V, A$, while the second to $\Gamma' = S, P$ or $T$. To guarantee the $z$-independence of \cref{eq:lept-me-forms} we need to further require $q_{\ell, \nu}$ to be of the form
\begin{equation}
(q_\nu)_{\dot\alpha \alpha} = | r_\nu ]_{\dot \alpha} \langle \nu |_\alpha~,  \quad (q_\ell)_{\dot\alpha \alpha} = | {\bf \ell}^{\bf L} ]_{\dot \alpha} \langle r_{\ell, {\bf L}} |_\alpha~, \ \ {\rm or} \ \ (q_\ell)_{\dot\alpha \alpha} = | r_{\ell}^{\bf L} ]_{\dot \alpha} \langle {\bf \ell}_{\bf L} |_\alpha~.
\end{equation}
Namely only  $| \nu ]$ and either $| {\bf \ell}^{\bf L}\rangle$ or $| {\bf \ell}^{\bf L}]$  are modified, such as the shifted spinors never enter the matrix element.
One can easily verify that the following deformations satisfy all the requirements: $q_\nu$ automatically satisfy $q_\nu^2=q_\nu \cdot p_\nu = 0$. Imposing the corresponding conditions $q_\ell^2= q_\ell \cdot p_\ell = 0$ for each  ansatz determines the form of $| r_{\ell}^{\bf L}\rangle$ and $| r_{\ell}^{\bf L}]$ respectively. For the case of $| r_\ell^{\bf L} ]$ one obtains 
\begin{equation}
| r_{\ell}^1 ] = -a c \big( | p_{\ell}^1 ] - c | p_{\ell}^2 ]\big)\,, \quad | r_{\ell}^2 ] = -a \big( | p_{\ell}^1 ] - c | p_{\ell}^2 ] \big), \\
\end{equation}
with $a$, $c$ undetermined constants. After some manipulations one arrives to writing $(q_\ell)_{\dot\alpha\alpha} \equiv | q_\ell]_{\dot\alpha} \langle q_{\ell}|_\alpha$ as
\begin{equation}
| q_\ell]\langle q_\ell| = a \big( | p_{\ell}^1 ] - c | p_{\ell}^2 ] \big)\big( \langle p_{\ell}^1 | - c \langle p_{\ell}^2 | \big)
\end{equation}
Writing $(q)_{\dot\alpha\alpha}$ as $ | q ]_{\dot\alpha} \langle q|_\alpha$ one can solve $(q_\nu + q_\ell - q)_{\dot\alpha\alpha} = 0$. This is always possible, given that the one has four degrees of freedom in choosing $| r_\nu] $ and $| q_\ell ]$. Explicitly, one can choose a basis for the spinors to be $|q]$, $|q']$ with $[q q']=0$ (and similarly their conjugates for the angled spinors). After projecting $|r_\nu]$, $|p_\nu\rangle$, $|q_\ell]$, $|q_\ell\rangle$ onto this basis, enforcing the condition above produces a system of 4 equations in 4 unknowns, which is easy to solve and show that it always has a solution. We spare the reader the algebraic details as they are not particularly illuminating.

\section{Alternative deformations \label{app:alt-def} }
It is useful to consider how our analysis might differ from other deformations of external momenta. In total, we have five particles participating in the reaction $B \rightarrow D \pi \ell \nu$. Therefore, beyond the choice of $p_{\ell}$, $p_\nu$, and $p_\pi$ discussed above, one can also consider alternative deformations. 

At first pass, an attractive option is to deform the residual momenta of the heavy particles,  $k\rightarrow k+zq$ and $k'\rightarrow k'+zq$, which does not alter $p_\pi$ or $p_{\ell\nu}$. Unfortunately, to reach physical factorization channels one requires $q \sim \mathcal{O}(M)$ at which point the power counting of the theory is broken and one is effectively deforming the velocity labels $v$ and $v'$. This has several technical complications and we do not discuss it further. 

A different choice is to deform the residual momentum of the initial $B$-meson, $k\rightarrow k+zq$, and the pion, $p_\pi \rightarrow p_\pi + zq$. This leaves the momentum injected by the current unchanged and allows one to pick off factorization channels in the $D$ system (but not the $B$ system).  The ``invisible'' factorization channel is encoded in a contribution to the boundary term. A similar deformation of $k'\rightarrow k'-zq$ and $p_\pi \rightarrow p_\pi + zq$ allows one to access factorization channels in the $B$ system (but not the $D$ system which is again encoded in the pole at infinity). A combination of the two approaches then allows one to obtain off-shell vertices for both systems but renders the analysis of the boundary term more complicated.

As a sketch of how this works let us consider the deformation of $k$. We require a $q$ that satisfies 
\begin{equation}
    q^2 =0 ~~~,~~~ k\cdot q = 0 ~~~~,~~~~ p_\pi \cdot q=0~. 
\end{equation}
The solutions of $z_\star'$ correspond to those obtained above. This is easily seen by noting that we have used the same deformation for the pion line and the final state $D$-meson is still on-shell. 

When performing the RPI transformation for the hard-current vertex we must now transform both $v$ and $v'$. We therefore have 
\begin{equation}
    \begin{split}
    w \rightarrow \tilde{w} &= (v-z_\star'q/M)\cdot (v'-z_\star'q/M' - p_\pi/M' ) \\
                            &= w -z_\star' v\cdot q/M' - z_\star'v'\cdot q/M' - v \cdot p_\pi/M'~.
    \end{split}
\end{equation}
After expanding the Isgur-Wise function, one of these terms pinches against the $D$-propagator and can be lumped into the contact terms at $\mathcal{O}(1/M)$. The contribution from $v'\cdot q$ (already encountered above) can be rewritten when combined with the propagator using $v'\cdot q/v\cdot q=M/M'$. The off-shell vertex correction that accompanies a pole in the low energy theory, (i.e., the term proportional to $v\cdot p_\pi/M$) is identical to that identified using the deformation chosen in the main text. This demonstrates how our procedure can identify off-shell vertex corrections that accompany poles, but not ``pinched'' contributions, which are ambiguous until the contour at $R_\infty$ is evaluated. This would correspond to a complete matching calculation in the EFT, however, this can be circumvented if Wilson coefficients can be fixed with low-energy data. 

When using either of the $k$ or $k'$ deformation, some of the factorization channels are hidden. These then appear in the contour at large-$z$ (i.e., the pole at infinity). By using a given deformation, the Feynman rule for an off-shell vertex in the effective theory can be inferred, independent of the other factorization channels in the problem. Using both the $k$ and $k'$ deformations one can construct all relevant off-shell vertices.

\section{BCFW Boundary term calculation}\label{app:infinity}

Here we derive in detail the contribution from the large-$z$ contour $\gamma$ in the BCFW factorization formula. The main observations determining our procedure is that (a) \hhchpt has a definite expansion in powers of $1/\Lambda_\chi$ and (b) $z$ enters in amplitudes only through $p_\pi(z)$ (as $v^{(\prime)}$ are undeformed by the complex shift and $\pln(z)$ never enters explicitly the expressions). 

The large-$z$ behavior is therefore controlled by the number of deformed pion momenta in the numerator of a term in the amplitude. This is bounded from above by the chiral order. To ameliorate the large-$z$ behavior it is therefore sufficient to take enough derivatives with respect to the deformed pion momentum $p_\pi(z)$ which we will denote $\hat p_\pi$ in the following. One can consequently write a Cauchy relation for the $k^{\rm th}$ derivative at $n^{\rm th}$ order in \hhchpt:
\begin{equation}
    \begin{split}\label{eq:Cauchy-deriv}
       \!\!\!\! \frac{1}{(2\pi\iu)} &\oint_\gamma \frac{\dd z'}{z'-z}  \frac{\partial^k {\cal A}^{( n )}}{\partial \hat p_\pi^{\mu_1}\ldots \partial \hat p_\pi^{\mu_k}}(z')  = \frac{\partial^k {\cal A}^{( n )}}  {\partial \hat p_\pi^{\mu_1}\ldots \partial \hat p_\pi^{\mu_k}}(z) \\ 
        &  +\sum_{i \in \textrm{poles}}  \Bigg\{ \bigg[{\rm Res}_1~\frac{1}{z'-z} \frac{\partial^k {\cal A}^{( n )}}{\partial \hat p_\pi^{\mu_1}\ldots \partial \hat p_\pi^{\mu_k}}(z') \bigg]_{z_i} + \bigg[{\rm Res}_2~\frac{1}{z'-z} \frac{\partial^k {\cal A}^{( n )}}{\partial \hat p_\pi^{\mu_1}\ldots \partial \hat p_\pi^{\mu_k}}(z')\bigg]_{z_i} \Bigg\} ~.
    \end{split}
 \end{equation}
By taking a sufficiently large number of derivatives we can gain control of the large radius behavior of the contour integral. For each $n$, there exists a $k_{\rm min}(n)$ for which the boundary term vanishes.

The above Cauchy relation determines the $k_{\rm min}^{\rm th}$ derivative of $\mathcal{A}^{(n)}$ in terms of on-shell poles. We can then proceed to determine all the $k^{\rm th}$ derivatives with $k< k_{\rm min}$ recursively. To guarantee a good behavior at large-$z$ one should apply the Cauchy relation to the suitably subtracted quantity:
\begin{equation}
    \frac{\partial^k }{\partial \hat p_\pi^{\mu_1}\ldots \partial \hat p_\pi^{\mu_k}} \left[{\cal A}^{( n )}(z) - \sum_{h = k+1}^{k_{\rm min}} \frac{1}{h!} p_{\pi}^{\mu_1}(z)\ldots \hat p_\pi^{\mu_h}(z) \frac{\partial^h {\cal A}^{( n )}}{\partial \hat p_\pi^{\mu_1}\ldots \partial \hat p_\pi^{\mu_h}}(z)\right]~,
\end{equation}
which is built using quantities computed at earlier steps of the recursion. Due to the subtraction, the boundary term does not receive any contribution from the large-$z$ growth of $p_\pi(z)$ from pole-like terms in the amplitude. Since in the case studied in this paper there are no spurious poles, the only non-vanishing contributions to the pole at infinity can come from contact terms at ${\cal O}(n)$ in the \hhchpt expansion. These have non-zero $k^{\rm th}$ derivatives with respect to $p_\pi(z)$, and can constructed using a series of Wilson coefficients and effective operators to parameterize the large-$z$ contour in its most general form.

As an example, if one applies this method to ${\cal A}(B \to D \ell \nu \pi)$, and expands the amplitude order by order in $1/\Lambda_\chi^n$ i.e., $\mathcal{A}= \mathcal{A}^{(0)} +  \mathcal{A}^{(1)} + \ldots$, one has, 

\begin{itemize}
\item ${\cal A}^{(0)}$: 

the leading order amplitude scales as $p_\pi(z)^1$, while the corresponding $1/M$ corrections as $p_\pi(z)^2$. Taking the second derivative with respect to $p_\pi(z)$ fixes the $1/M$ corrections, then the first derivative fixes the ${\cal A}^{(0)}$. No contact terms can be written at $\mathcal{O}(1/\Lambda_\chi^0)$ with non-zero first and second derivatives with respect to $p_\pi$, so the boundary terms vanish. Since RPI invariance fixes the relative structure between the terms linear and quadratic in $p_\pi(z)$, in principle taking the second derivative will be sufficient to fix the whole amplitude, but we will not use this shortcut to illustrate the process for more general situations.
Explicitly, we can use \cref{vicin-pole-noprime,vicin-pole-prime} for the near-pole form of the amplitude and further write the numerators as:
\begin{align}
\mathcal{A}^{(s)}_L \mathcal{A}^{(h)}_R(z) &= p_\pi^\mu(z) \Bigg[v^\nu -\frac{p_\pi^\nu(z)}{M}\Bigg]B_{\mu\nu} ~,\\
\mathcal{A}^{(h)}_L \mathcal{A}^{(s)}_R(z) &= p_\pi^\mu(z) \Bigg[v^{\prime\nu} +\frac{p_\pi^{\prime\nu}(z)}{M'}\Bigg]B'_{\mu\nu}~,
\end{align}
where we have used an RPI invariant form for the $1/M^{(')}$ terms. Using \cref{eq:Cauchy-deriv} on the second derivative of $\mathcal{A}^{(0)}$ and on the first derivative of 
\begin{equation}
    \mathcal{A}^{(0,\textrm{sub})}(z) \equiv \mathcal{A}^{(0)}(z) - \frac{ p_\pi^\mu(z) p_\pi^\nu(z)}{2} \frac{\partial^2 \mathcal{A}^{(0)}}{\partial \hat p_\pi^\mu \partial \hat p_\pi^\nu}(z)~, \label{eq:amp-subtracted}
\end{equation}
 one obtains
\begin{align}
\frac{\partial^2 {\cal A}^{( 0 )}}{\partial \hat p_\pi^{\mu_1} \partial \hat p_\pi^{\mu_2}}(z)  & =  \frac{ B_{\{\mu_1,\mu_2\}}}{(z- z_\star) M v \cdot q}  + \frac{ B'_{\{\mu_1,\mu_2\}}}{(z- z'_\star) M' v' \cdot q} \\
\frac{\partial {\cal A}^{( 0, \textrm{sub} )}}{\partial \hat p_\pi^{\mu} }(z) &= \begin{multlined}[t]
- \frac{1}{2(z- z_\star) v \cdot q} \Big[ B_{\mu\nu} \Big(v^\nu - \frac{\hat p_\pi^\nu}{M}\Big) - \frac{\hat p_\pi^\nu}{M} B_{\nu\mu}\Big] \\
 + \frac{1}{2(z-z'_\star) v' \cdot q} \Big[B'_{\mu\nu} \Big(v^{\prime\nu} + \frac{\hat p_\pi^\nu}{M'}\Big) + \frac{\hat p_\pi^\nu}{M'} B'_{\nu\mu}\Big] \\
 + \frac{1}{((z-z_\star) v \cdot q)^2}\frac{m_\pi^2- (\Delta M)^2}{4 M} B_{\mu\nu} v^{\nu} \\
  + \frac{1}{((z-z'_\star) v' \cdot q)^2}\frac{m_\pi^2- (\Delta M')^2}{4 M'} B'_{\mu\nu}v^{\prime\nu}~.
\end{multlined}
\end{align}
where $B^{(')}_{\{\mu,\nu\}} = \big( B^{(')}_{\mu\nu} + B^{(')}_{\nu\mu} \big)/2$ and in the second derivative we have dropped the double pole contribution as it is of ${\cal O}(1/M^{(\prime)2})$. 
The large-$z$ behavior of ${\cal A}^{(0)}(z)/z$ is
\begin{equation}\label{eq:C-gamma-0}
    \begin{split}
        \frac{{\cal A}^{(0)}(z)}{z} & \sim \frac{p_\pi^\mu(z)}{z} \frac{\partial {\cal A}^{( 0,\textrm{sub} )}}{\partial \hat p_\pi^{\mu} }(z)+\frac{p_\pi^\mu(z) p_\pi^\nu(z)}{2z} \frac{\partial^2 {\cal A}^{( 0 )}}{\partial \hat p_\pi^{\mu} \partial \hat p_\pi^{\nu}}(z) \\
        & \sim 
        \frac{q^\mu q^\nu}{2} \Bigg[\frac{B_{\mu\nu}}{M v \cdot q} + \frac{B'_{\mu\nu}}{M' v' \cdot q}\Bigg] \\
        &\hspace{0.1\linewidth}+ \frac{1}{2z}\Bigg[ \frac{B_{\mu\nu} q^\mu}{v \cdot q} \Big(v^\nu - \frac{p_\pi^{\nu} -z_\star q^\nu}{M}\Big) -\frac{B_{\mu\nu} q^\nu p_\pi^\mu}{ M v\cdot q} \\
        &\hspace{0.25\linewidth}-\frac{B'_{\mu\nu} q^\mu}{v' \cdot q} \Big(v^{\prime\nu} + \frac{p_\pi^{\nu} -z'_\star q^\nu}{M'}\Big) -\frac{B'_{\mu\nu} q^\nu p_\pi^\mu}{ M' v'\cdot q}\Bigg]\\
        &\hspace{0.5\linewidth}+ \mathcal{O}\Big(\frac{1}{z^2}\Big)~.
    \end{split}
\end{equation}
The term multiplying the $1/z$ term is nothing else than $C_\gamma^{(0)}$ and can be written as
\begin{equation}
    C_\gamma^{(0)} = \Bigg[\frac{\mathcal{A}^{(h)}_L \mathcal{A}^{(s)}_R(z'_\star)-\mathcal{A}^{(h)}_L \mathcal{A}^{(s)}_R(0)}{2z'_\star v' \cdot q} -\frac{\mathcal{A}^{(s)}_L \mathcal{A}^{(h)}_R(z_\star)-\mathcal{A}^{(s)}_L \mathcal{A}^{(h)}_R(0)}{2z_\star v \cdot q}\Bigg]~.\label{eq:Cgamma-leading}
\end{equation}

\item $\mathcal{A}^{(1)}$: 
the subleading power amplitude scales as $p_\pi(z)^2/\Lambda_\chi$, with the corresponding $1/M$ corrections being proportional to $p_\pi(z)^3$. Ignoring terms of ${\cal O}(1/M\Lambda_\chi)$ one can take the second derivative of ${\cal A}^{(1)}$. Similarly to what was done for ${\cal A}^{(0)}$, this will fix the boundary contribution in terms of pole terms at ${\cal O}(1/\Lambda_\chi)$, which originate from factorization channels where either the left or the right amplitude is taken at next to leading power. No contact terms are contributing to the left-hand side of \cref{eq:Cauchy-deriv} with $k=2$.  One then considers the subtracted amplitude $\mathcal{A}^{(1,\textrm{sub})}(z) $, defined similarly to \cref{eq:amp-subtracted}, and applies the Cauchy formula to its first derivative. In this case, however, the contour at large $z$ is non-zero as one can write contact terms proportional to $p_\pi(z)/\Lambda_\chi$ to the amplitude. Therefore one can parameterize the left-hand side of \cref{eq:Cauchy-deriv} with the first derivative of the most general set of contact operators contributing to ${\cal A}^{(1)}$:
\begin{equation}
\begin{split}
 \frac{1}{2\pi i} \oint_\gamma \frac{\dd z'}{z'-z}  \frac{\partial {\cal A}^{( 1 )}}{\partial \hat p_\pi^{\mu}}(z') & =  \frac{f_{1}(v \cdot v')}{\Lambda_\chi} \mel{D_{v'}(0)}{{\rm Tr}\big[~ \overline{H}'_{v'} \Gamma H_v \gamma^5 \slashed{\Api}'~\! (\hat p_\pi) \big]}{\Bbar_v(0)} \\ 
 &\hspace{0.025\linewidth}+ \frac{f_2(v \cdot v')}{\Lambda_\chi} \mel{D_{v'}(0)}{{\rm Tr}\big[~ \overline{H}'_{v'} \Gamma H_v \gamma^5 (v+v')\cdot \Api' ~\! (\hat p_\pi) \big]}{\Bbar_v(0)} \\
 &\hspace{0.025\linewidth}+ \frac{f_3(v \cdot v')}{\Lambda_\chi} \mel{D_{v'}(0)}{{\rm Tr}\big[~ \overline{H}'_{v'} \Gamma H_v \gamma^5 (v-v')\cdot \Api' ~\! (\hat p_\pi) \big]}~,{\Bbar_v(0)}\label{eq:cgamma-loc-der}
\end{split}
\end{equation}
where $\Api'(\hat p_\pi) = (\partial \Api / \partial \hat p_\pi^\mu)$. The result $C_\gamma^{(1)}$ will be the sum of the pole subtraction terms $C_\gamma^{(1,{\rm pole})}$ analogous to 
\cref{eq:Cgamma-leading} for leading power  and the contact term contribution $C_\gamma^{(1,{\rm local})}$ from contracting the right-hand side of \cref{eq:cgamma-loc-der} with $p_\pi$.

\end{itemize}
\makeatletter
\let\save@section\section
\renewcommand{\section}[1]{
	\save@section{#1}
	\addtocontents{toc}{\protect\vspace*{-10pt}}
}
\makeatother

\addtocontents{toc}{\protect\vspace*{10pt}}
\bibliography{refs}

\end{document}